\documentclass[12pt,journal,draftcls,letterpaper,onecolumn]{IEEEtran}
\usepackage{amsmath,cite,graphicx,pstricks,textcomp,txfonts,url}
\newcommand{\bF}{\textsf{\textbf{F}}}
\newcommand{\bP}{\textsf{\textbf{P}}}
\newcommand{\bt}{\mathbf{t}}
\newcommand{\bx}{\mathbf{x}}
\begin{document}
\title{The Context Sensitivity Problem in Biological Sequence Segmentation}
\author{Siew-Ann Cheong,\thanks{S.-A. Cheong completed this work as a
Postdoctoral Associate with the Cornell Theory Center, Cornell
University, Ithaca, NY 14853.  He is presently an Assistant Professor
of Physics and Applied Physics with the School of Physical and
Mathematical Sciences, Nanyang Technological University, 21 Nanyang
Link, Singapore 637371, Republic of Singapore.  Email:
cheongsa@ntu.edu.sg.}
Paul Stodghill,\thanks{P. Stodghill, D. J. Schneider and S. W. 
Cartinhour are with the USDA Agricultural Research Service, Ithaca, NY 
14853.  Email: ps27@cornell.edu, djs30@cornell.edu, 
sc167@cornell.edu.} 
David J. Schneider,
Samuel W. Cartinhour, and 
Christopher R. Myers\thanks{C. R. Myers is with the Center for
Advanced Computing, Cornell University, Ithaca, NY 14853.  Email:
myers@tc.cornell.edu.} }
\maketitle

\begin{abstract}

In this paper, we describe the context sensitivity problem encountered
in partitioning a heterogeneous biological sequence into statistically
homogeneous segments.  After showing signatures of the problem in the
bacterial genomes of \emph{Escherichia coli} K-12 MG1655 and
\emph{Pseudomonas syringae} DC3000, when these are segmented using two
entropic segmentation schemes, we clarify the contextual origins of
these signatures through mean-field analyses of the segmentation
schemes.  Finally, we explain why we believe all sequence segmentation
schems are plagued by the context sensitivity problem.

\end{abstract}

\section{Introduction}

Biological sequences are statistically heterogeneous, in the sense
that local compositions and correlations in different regions of the
sequences can be very different from one another.  They must therefore
treated as collections of statistically stationary segments (or
\emph{domains}), to be discovered by the various segmentation schemes
found in the literature (see review by Braun and M\"uller
\cite{Braun1998StatisticalScience13p142}, and list of references in
Ref.~\citeonline{Cheong2007IRJSSS}).  Typically, these segmentation
schemes are tested on (i) artificial sequences composed of a small
number of segments, (ii) control sequences obtained by concatenating
known coding and noncoding regions, or (iii) control sequences
obtained by concatenating sequences from chromosomes know to be
statistically distinct.  They are then applied on a few better
characterized genomic sequences, and compared against each other, to
show general agreement, but also to demonstrate better sensitivity in
delineating certain genomic features.  To the best of our knowledge,
there are no studies reporting a full and detailed comparison of the
segmentation of a sequence against its distribution of carefully
curated gene calls.  There are also no studies comparing the
segmentations of closely related genomes.  In such sequences, there
are homologous stretches, interrupted by lineage specific regions, and
the natural question is whether homologous regions in different
genomes will be segmented in exactly the same way by the same
segmentation scheme.

In this paper, we answer this question, without comparing the
segmentation of homologous regions.  Instead, through careful
observations of how segment boundaries, or \emph{domain walls}, are
discovered by two different entropic segmentation schemes, we realized
that a subsequence can be segmented differently by the same scheme, if
it is part of two different full sequences.  We call this dependence
of a segmentation on the detailed arrangement of segments the
\emph{context sensitivity problem}.  In
Sec.~\ref{section:contextsensitivityprobleminrealgenomes}, we will
describe how the context sensitivity problem manifests itself in real
genomes, when these are segmented using a sliding-window entropic
segmentation scheme, which examines local contexts in the sequences,
versus segmentation using a recursive entropic segmentation scheme,
which examines the global contexts of the sequences.  We then show how
the context sensitivity problem prevents us from coarse graining by
using larger window sizes, stopping recursive segmentation earlier, or
by simply removing weak domain walls from a fine-scale segmentation.
We follow up in Sec.~\ref{section:meanfieldanalysis} with a mean-field
analysis of the local and global context sensitivity problems, showing
how the positions and strengths of domain walls, and order in which
these are discovered, are affected by these contexts.  In particular,
we identify repetitive sequences as the worst case scenario to
encounter during segmentation.  Finally, in
Sec.~\ref{section:conclusions}, we summarize and discuss the impacts
of our findings, and explain why we believe the context sensitivity
problem plagues \emph{all} segmentation schemes.

\section{Context Sensitivity Problem in Real Bacterial Genomes}
\label{section:contextsensitivityprobleminrealgenomes}

In this section, we investigate the manifestations of the context
sensitivity problem in two real bacterial genomes, those of
\emph{Escherichia coli} K-12 MG1655 and \emph{Pseudomonas syringae}
DC3000, when these are segmented using two entropic segmentation
schemes.  The first entropic segmentation scheme, based on statistics
comparison of a pair of sliding windows, is sensitive to the local
context of segments within the pair of sliding windows, and we shall
show in Sec.~\ref{subsection:pairedslidingwindows} that the positions
and strengths of domain walls discovered by the scheme depends
sensitively on the window size.  The second entropic segmentation
scheme is recursive in nature, adding new domain walls at each stage
of the recursion.  We shall show in
Sec.~\ref{subsection:recursivegenome} that this scheme is sensitive to
the global context of segments within the sequence, and that domain
walls are not discovered according to their true strengths.  In
Sec.~\ref{subsection:bottomupsegmentationhistory}, we show that there
is no statistically consistent way to coarse grain a segmentation by
removing the weakest domain walls, and agglomerating adjacent
segments.

\subsection{Paired Sliding Windows Segmentation Scheme}
\label{subsection:pairedslidingwindows}

Using the paired sliding windows segmentation scheme described in
App.~\ref{section:pairedslidingwindowssegmentationscheme}, the number
$M$ of order-$K$ Markov-chain segments discovered depends on the size
$n$ of the windows used, as shown in Table \ref{table:K0n} for
\emph{E. coli} K-12 MG1655.  Because $M$ decreases as $n$ is
increased, we are tempted to think that we can change the granularity
of the segmental description of a sequence by tuning $n$, such that
there are more and shorter segments when $n$ is made smaller, while
there are fewer and longer segments when $n$ is made larger.  Thus, as
$n$ is increased, we expect groups of closely spaced domain walls to
be merged as the short segments they demarcate are agglomerated, and
be replaced by a peak close to the position of the strongest peak.

\begin{table}[hbtp]
\centering
\caption{Number of $K = 0$ domain walls in the \emph{E. coli} K-12
MG1655 genome ($N = 4639675$ bp), obtained using the paired sliding
window segmentation scheme for different window sizes $1000 \leq n
\leq 5000$.}
\label{table:K0n}
\vskip .5\baselineskip
\begin{tabular}{|c|c|c|c|c|c|}
\hline
$n$ & 1000 & 2000 & 3000 & 4000 & 5000 \\
\hline
$M$ & 2781 & 1414 & 952 & 721 & 577 \\
\hline
\end{tabular}
\end{table}
 
\begin{figure}[hbtp]
\centering
\includegraphics[scale=0.5,clip=true]%
	{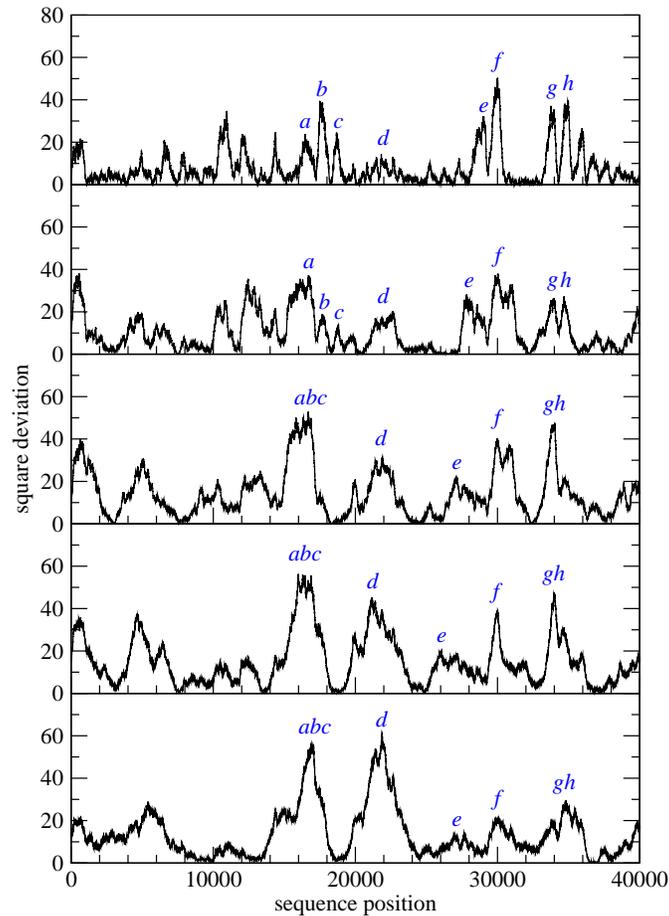}
\caption{The $K = 0$ square deviation spectra in the region $(0,
40000)$ of the \emph{E. coli} K-12 MG1655 genome, obtained using the paired
sliding window segmentation scheme with window sizes (top to bottom)
$n = 1000$, 2000, 3000, 4000, and 5000.}
\label{figure:EcoliK12qrK0n1kn2kn3kn4kn5ki0i40k}
\end{figure}

\begin{figure}[hbtp]
\centering
\includegraphics[scale=0.5,clip=true]%
	{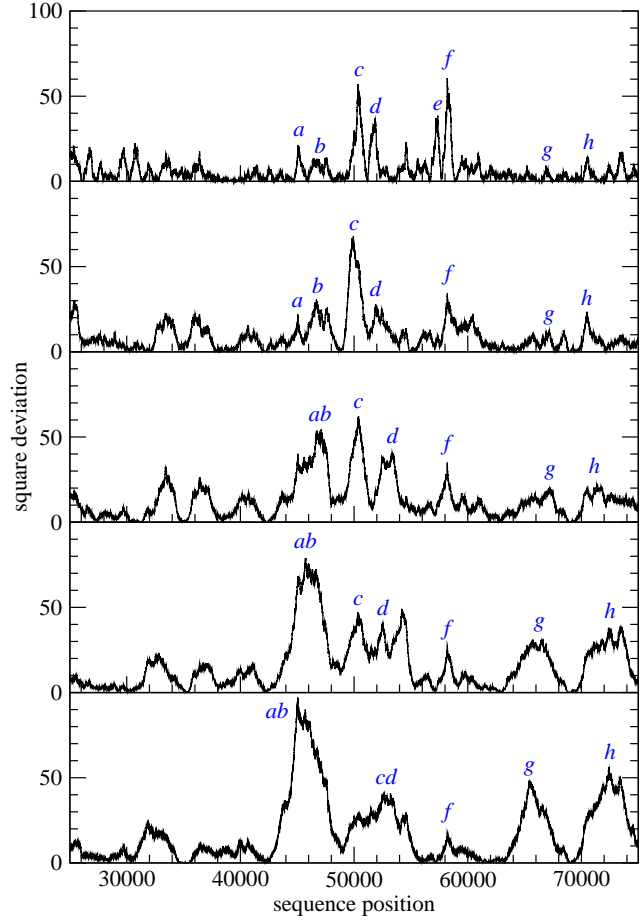}
\caption{The $K = 0$ square deviation spectra in the region $(25000,
75000)$ of the \emph{P. syringae} DC3000 genome, obtained using the
paired sliding window segmentation scheme with window sizes (top to
bottom) $n = 1000$, 2000, 3000, 4000, and 5000.}
\label{figure:PsyringaeqrK0n1kn2kn3kn4kn5ki25ki75k}
\end{figure}

Indeed, we do find this expected merging of proximal domain walls in
Fig.~\ref{figure:EcoliK12qrK0n1kn2kn3kn4kn5ki0i40k} and
Fig.~\ref{figure:PsyringaeqrK0n1kn2kn3kn4kn5ki25ki75k}, which shows
the square deviation spectra for the $(0, 40000)$ region of the
\emph{E.  coli} K-12 MG1655 genome and the $(25000, 75000)$ region of
the \emph{P. syringae} DC3000 genome respectively.  In the $(0,
40000)$ region of the \emph{E. coli} K-12 MG1655 genome shown in
Fig.~\ref{figure:EcoliK12qrK0n1kn2kn3kn4kn5ki0i40k}, we find the group
of domain walls, $i_a \approx 16500$, $i_b \approx 17500$, and $i_c
\approx 18700$, and the pair of domain walls, $i_g \approx 33800$ and
$i_h \approx 35000$, which are distinct in the $n = 1000$ square
deviation spectrum, merging into the domain walls $i_{abc}$ and
$i_{gh}$ in the $n \geq 3000$ square deviation spectra.  In the
$(25000, 75000)$ region of the \emph{P. syringae} DC3000 genome shown
in Fig.~\ref{figure:PsyringaeqrK0n1kn2kn3kn4kn5ki25ki75k}, we find the
pair of domain walls, $j_a \approx 45000$ and $j_b \approx 46600$, and
the pair of domain walls, $j_c \approx 50400$ and $j_d \approx 51800$,
which are distinct in the $n = 1000$ square deviation spectrum,
merging into the domain walls $j_{ab}$ and $j_{cd}$ in the $n \geq
3000$ and $n = 5000$ square deviation spectra respectively.

However, we also find unexpected changes in the relative strengths of
the domain walls, as $n$ is increased.  In the $(0, 40000)$ region of
the \emph{E. coli} K-12 MG1655 genome shown in
Fig.~\ref{figure:EcoliK12qrK0n1kn2kn3kn4kn5ki0i40k}, we find that $i_d
\approx 21800$, which appears as a broad, weak, and noisy bump in the
$n = 1000$ square deviation spectrum, becoming stronger and more
defined as $n$ is increased, and finally becomes as strong as the
domain wall $i_{abc}$ in the $n = 5000$ square deviation spectrum.  In
this region of the \emph{E. coli} K-12 MG1655 genome, we also find
that the domain walls $i_b \approx 17500$ and $i_f \approx 30000$ are
equally strong in the $n = 1000$ square deviation spectrum, but as $n$
is increased, $i_b$ becomes stronger while $i_f$ becomes weaker.  In
the $(25000, 75000)$ region of the \emph{P. syringae} DC3000 genome
shown in Fig.~\ref{figure:PsyringaeqrK0n1kn2kn3kn4kn5ki25ki75k}, we
find that the domain walls $j_c \approx 50400$ and $j_f \approx 58200$
are equally strong, and also the domain walls $j_d \approx 51800$ and $j_e
\approx 57300$ are equally strong, in the $n = 1000$ square deviation
spectrum.  However, as $n$ is increased, $j_c$ becomes stronger than
$j_f$, while $j_d$ becomes stronger than $j_e$.  More importantly,
all these domain walls --- the strongest in this $(25000, 75000)$
region of the $n = 1000$ square deviation spectrum --- become weaker
as $n$ is increased, to be superseded by the domain walls $j_{ab}
\approx 45000$, $j_g \approx 65400$ and $j_h \approx 72400$, which
become stronger as $n$ is increased.  As it turned out, $(j_c, j_f)$
overlaps significantly with the interval interval $(50000, 59000)$,
which incorporates three lineage-specific regions (LSRs 5, 6, and 7,
all of which virulence related) identified by Joardar \emph{et al}
\cite{Joardar2005MolPlantPathol6p53}.  It is therefore biologically
significant that $j_c$ and $j_f$ are strong domain walls in the $n =
1000$ square deviation spectrum.  On the other hand, it is not clear
what kind of biological meaning we can attach to $j_{ab}$, $j_g$, and
$j_h$ being the strongest domain walls in the $n = 5000$ square
deviation spectrum.  

\begin{table}[htbp]
\centering
\caption{Positions of strong domain walls in the $(0, 40000)$ region
of the \emph{E. coli} K-12 MG1655 genome and the $(25000, 75000)$
region of the \emph{P. syringae} DC3000 genome, determined after match
filtering the square deviation spectra obtained using the paired
sliding window segmentation scheme with window sizes $n = 3000, 4000,
5000$.}
\label{table:slidingwindowshifts}
\vskip .5\baselineskip
\begin{tabular}{|c|c|c|c|c|c|c|}\hline
& \multicolumn{3}{c|}{\emph{E. coli} K-12 MG1655} &
\multicolumn{3}{c|}{\emph{P. syringae} DC3000} \\ \hline
$n$ & $i_{abc}$ & $i_d$ & $i_h$ & $j_{ab}$ & $j_g$ & $j_h$ \\ \hline
3000 & 16200 & 21800 & 34100 & 46600 & 66600 & 71500 \\
4000 & 16300 & 21700 & 34400 & 45900 & 65900 & 72500 \\
5000 & 16100 & 22100 & 34700 & 45700 & 65500 & 72500 \\ \hline
\end{tabular}
\end{table}

There is another, more subtle, effect that increasing the size of the
sliding windows has on the domain walls: their positions, as
determined from peaks in the square deviation spectrum after match
filtering, are shifted.  The shifting positions of some of the strong
domain walls in the $(0, 40000)$ region of the \emph{E. coli} K-12
MG1655 genome and the $(25000, 75000)$ region of the \emph{P.
syringae} DC3000 genome are shown in Table
\ref{table:slidingwindowshifts}.  In general, the positions and
strengths of domain walls can change when the window size used in the
paired sliding windows segmentation scheme is changed, because windows
of different sizes examine different local contexts.  As a result of
this local context sensitivity, whose nature we will illustrate using
a mean-field picture in
Sec.~\ref{subsection:meanfieldwindowedspectrum}, the sets of strong
domain walls determined using two different window sizes $n$ and $n' >
n$ are different.  If $n$ and $n'$ are sufficiently different, the
sets of strong domain walls, i.e.~those stronger than a specified
cutoff, may have very little in common.  Therefore, we cannot think of
the segmentation obtained at window size $n'$ as the coarse grained
version of the segmentation obtained at window size $n$.

\subsection{Optimized Recursive Jensen-Shannon Segmentation Scheme}
\label{subsection:recursivegenome}

\begin{figure}[htbp]
\centering
\includegraphics[scale=0.8]{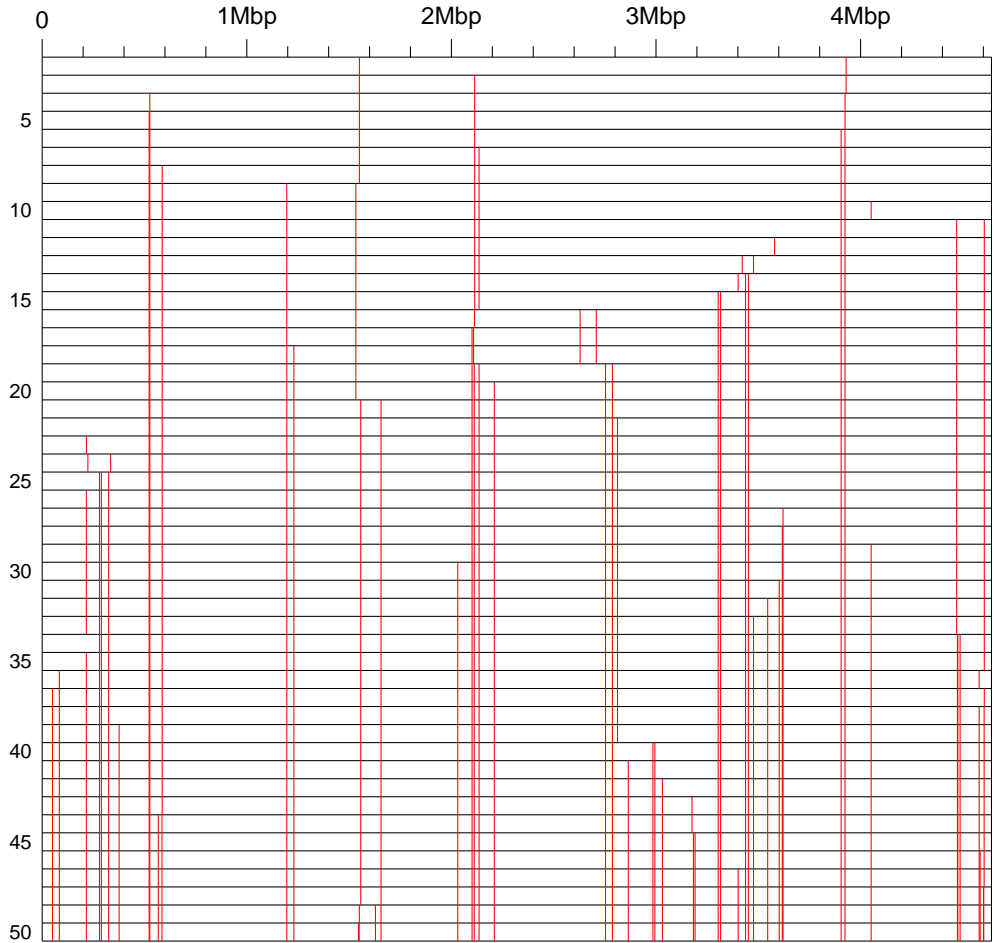}
\caption{Series of optimized recursive Jensen-Shannon segmentations of
the \emph{E. coli} K-12 MG1655 genome, for (top to bottom) $2 \leq M
\leq 50$ domain walls.  The two stable domain walls that appear in the
$M = 2$ optimized segmentation are close to the replication origin and
replication terminus.}
\label{figure:hierarchyofrecursivesegmentations}
\end{figure}

\begin{figure}[htbp]
\centering
\includegraphics[scale=0.8]{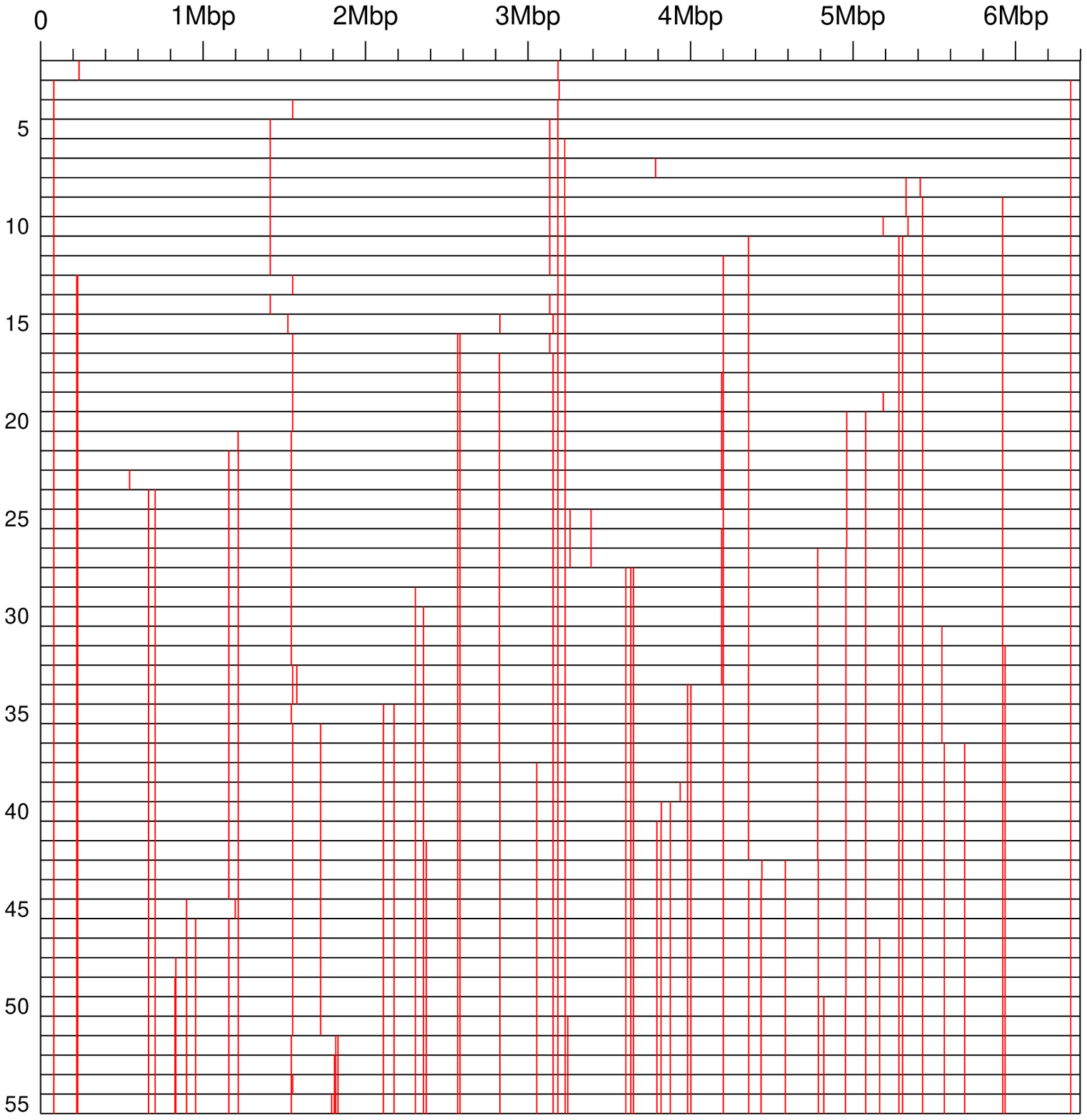}
\caption{Series of optimized recursive Jensen-Shannon segmentations of
the \emph{P. syringae} DC3000 genome, for (top to bottom) $2 \leq M
\leq 55$ domain walls.  Compared to the \emph{E. coli} K-12 MG1655
genome, there are perceptibly more unstable domain walls in the
\emph{P. syringae} DC3000 genome.}
\label{figure:pstm2m50}
\end{figure}

Using the optimized recursive Jensen-Shannon segmentation scheme
described in Ref.~\citeonline{Cheong2007IRJSSS}, we obtained one
series of segmentations each for \emph{E. coli} K-12 MG1655 and
\emph{P. syringae} DC3000, shown in
Fig.~\ref{figure:hierarchyofrecursivesegmentations} and
Fig.~\ref{figure:pstm2m50} respectively.  Two features are
particularly striking about these plots.  First, there exist domain
walls stable with respect to segmentation optimization.  These
\emph{stable domain walls} remain close to where they were first
discovered by the optimized recursive segmentation scheme.  Second,
there are \emph{unstable domain walls} that get shifted by as much as
10\% of the total length of the genome when a new domain wall is
introduced.  For example, in
Fig.~\ref{figure:hierarchyofrecursivesegmentations} for the \emph{E.
coli} K-12 MG1655 genome, we find the domain wall $i_{10} = 4051637$
in the optimized segmentation with $M = 10$ domain walls shifted to
$i_{10} = 4469701$ in the optimized segmentation with $M = 11$ domain
walls ($\delta i_{10} = +418064$), and also the domain wall $i_7 =
2135183$ in the optimized segmentation with $M = 15$ domain walls
shifted to $i_7 = 2629043$ in the optimized segmentation with $M = 16$
domain walls ($\delta i_7 = +493860$).  Based on the observation that
some unstable domain walls are discovered, lost, later rediscovered
and become stable, we suggested in Ref.~\citeonline{Cheong2007IRJSSS}
that for a given segmentation with $M$ domain walls, stable domain
walls are statistically more significant than unstable domain walls,
while stable domain walls discovered earlier are more significant than
stable domain walls discovered later in the optimized recursive
segmentation.

From Fig.~\ref{figure:hierarchyofrecursivesegmentations} and
Fig.~\ref{figure:pstm2m50}, we also find that the \emph{E. coli} K-12
MG1655 and \emph{P. syringae} DC3000 genomes have very different
segmental textures.  At this coarse scale ($M \sim 50$ segments), we
find many short segments, many long segments, but few segments of
intermediate lengths in the \emph{E. coli} K-12 MG1655 genome.  In
contrast, at the same granularity, the \emph{P.  syringae} DC3000
genome contains many short segments, many segments of intermediate
lengths, but few long segments.  We believe these segmental textures
are consistent with the different evolutionary trajectories of the two
bacteria.  \emph{E. coli} K-12 MG1655, which resides in the highly
stable human gut environment, has a more stable genome containing
fewer large-scale rearrangements which appear to be confined to
hotspots within the $(2600000, 3600000)$ region.  The genome of
\emph{P.  syringae} DC3000, on the other hand, has apparently
undergone many more large-scale rearrangements as its lineage
responded to multiple evolutionary challenges living in the hostile
soil environment.

We find many more large shifts in the optimized domain wall positions
in \emph{P. syringae} DC3000 compared to \emph{E. coli} K-12 MG1655,
because of the more varied context of the \emph{P.  syringae} DC3000
genome.  However, large shifts in the optimized domain wall positions
arise generically in all bacterial genomes, because of the sensitivity
of optimized domain wall positions to the contexts they are restricted
to.  In Sec.~\ref{subsection:meanfieldwindowlessspectrum}, we will
illustrate using a mean-field picture how the recursive segmentation
scheme decides where to subdivide a segment, i.e.~add a new domain
wall, after examining the global context within the segment.  We then
show how this global context changes when the segment is reduced or
enlarged during segmentation optimization, which can then cause a
large shift in the position of the new domain wall.  Because of this
\emph{global context sensitivity}, we find in
Fig.~\ref{figure:pstm2m50} a large shift of the domain wall $j_9 =
1723734$, which is stable when there are $36 \leq M \leq 51$ optimized
domain walls in the segmentation, to its new position $j_9 = 1818461$
($\delta j_9 = +94727$) when one more optimized domain wall is added.
We say that a domain wall is \emph{stable at scale $M$} if it is only
slightly shifted, or not at all, within the optimized segmentations
with between $M - \delta M$ and $M + \delta M$ domain walls, where
$\delta M \ll M$.  Given a series of recursively determined optimized
segmentations, we know which domain walls in an optimized segmentation
containing $M$ domain walls are stable at scale $M$, and which domain
walls in an optimized segmentation containing $M' > M$ domain walls
are stable at scale $M'$.  However, these two sets of stable domain
walls can disagree significantly because of the recursive segmentation
scheme's sensitivity to global contexts.  Again, we cannot think of
the optimized segmentation containing $M$ domain walls as a coarse
grained version of the optimized segmentation containing $M'$ domain
walls.

\subsection{Coarse-Graining by Removing Domain Walls}
\label{subsection:bottomupsegmentationhistory}

In Sec.~\ref{subsection:pairedslidingwindows}, we saw the difficulties
in coarse graining the segmental description of a bacterial genome by
using larger window sizes, due to the paired sliding windows
segmentation scheme's sensitivity to local context.  We have also seen
in Sec.~\ref{subsection:recursivegenome} a different set of problems
associated with coarse graining by stopping the optimized recursive
Jensen-Shannon segmentation earlier, due this time to the scheme's
sensitivity to global context.  Another way to do coarse graining
would be to start from a fine segmentation, determined using a paired
sliding window segmentation scheme with small window size, or properly
terminated recursive segmentation scheme, and then remove the weakest
domain walls.  Our goal is to agglomerate shorter, weakly distinct
segments into longer, more strongly distinct segments.  Although this
sounds like the recursive segmentation scheme playbacked in reverse,
there are subtle differences: in the recursive segmentation scheme,
strong domain walls may be discovered after weak ones are discovered,
so our hope with this coarse graining scheme is that we target weak
domain walls after `all' domain walls are discovered.

\begin{figure}[htbp]
\centering
\includegraphics[scale=0.45,clip=true]{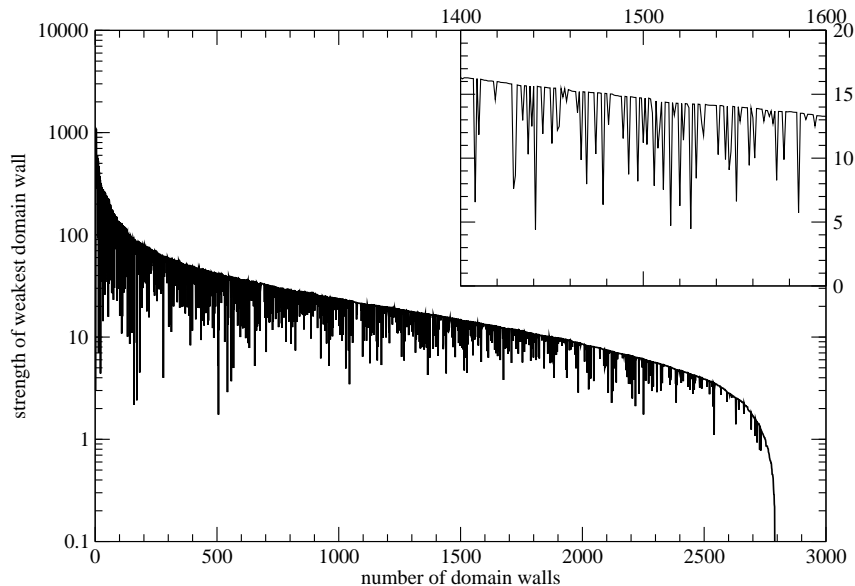}
\caption{Bottom-up segmentation history for \emph{E. coli} K-12 MG1655
derived from the initial $(K = 0, n = 1000)$ paired sliding windows
segmentation containing $M = 2781$ domain walls.  (Inset) Bottom-up
segmentation history from $M = 1600$ domain walls remaining to $M =
1400$ domain walls remaining, showing the fine structure of dips below
the smooth envelope.}
\label{figure:EcoliK12qtopsegKK0n1000}
\end{figure}

Like recursive segmentation, there are many detail variations on the
implementation of such a coarse graining scheme.  The first thing we
do is to select a cutoff strength $\Delta^*$, which we can think of as
a knob we tune to get a desired granularity for our description of the
genome: we keep a large number of domain walls if $\Delta^*$ is small,
and keep a small number of domain walls if $\Delta^*$ is large.  After
selecting $\Delta^*$, we can then remove all domain walls weaker than
$\Delta^*$ in one fell swoop, or remove them progressively, starting
from the weakest domain walls.  However we decide to remove domain
walls weaker than $\Delta^*$, the strengths of the remaining domain
walls must be re-evaluated after some have been removed from the
segmentation.  This is done by re-estimating the maximum-likelihood
transition probabilities, and using them to compute the Jensen-Shannon
divergences between successive coarse-grained segments, which are the
strengths of our remaining domain walls.  For the purpose of
benchmarking, we start from the $(K = 0, n = 1000)$ paired sliding
windows segmentation containing $M = 2781$ domain walls for the
\emph{E.  coli} K-12 MG1655 genome, and remove the weakest domain wall
each time to generate a \emph{bottom-up segmentation history}, shown
in Fig.~\ref{figure:EcoliK12qtopsegKK0n1000}.  As we can see, the
strength of the weakest domain wall as a function of the number of
domain wall remaining consists of a smooth envelope, and dips below
this envelope.  We distinguish between sharp dips, which are the
signatures of what we called \emph{tunneling events}, and broad dips,
which are the signatures of what we called \emph{cascade events}.  

\begin{figure}[hbt]
\centering
\includegraphics[scale=0.8,clip=true]{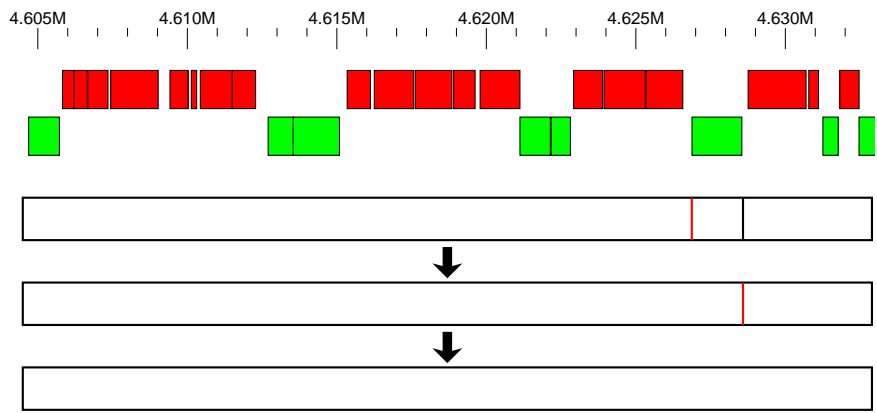}
\caption{A tunneling event occuring between $M = 1586$ and $M = 1584$
domain walls remaining in the bottom-up segmentation history of
\emph{E. coli} K-12 MG1655 ($N = 4639675$ bp), starting from the $(K =
0, n = 1000)$ initial segmentation containing $M = 2791$ domain walls.
Three segments in the $(4604497, 4632896)$ region of the genome are
shown.  The short segment involved in this tunneling event consists of
the single gene \emph{yjjX} on the negative strand (green), flanked by
two segments consisting of genes found predominantly on the positive
strand (red).  At each stage of the bottom-up segmentation history,
the domain wall removed is highlighted in red.}
\label{figure:NC000913topsegKL1586}
\end{figure}

\begin{figure}[hbt]
\centering
\includegraphics[scale=0.8,clip=true]{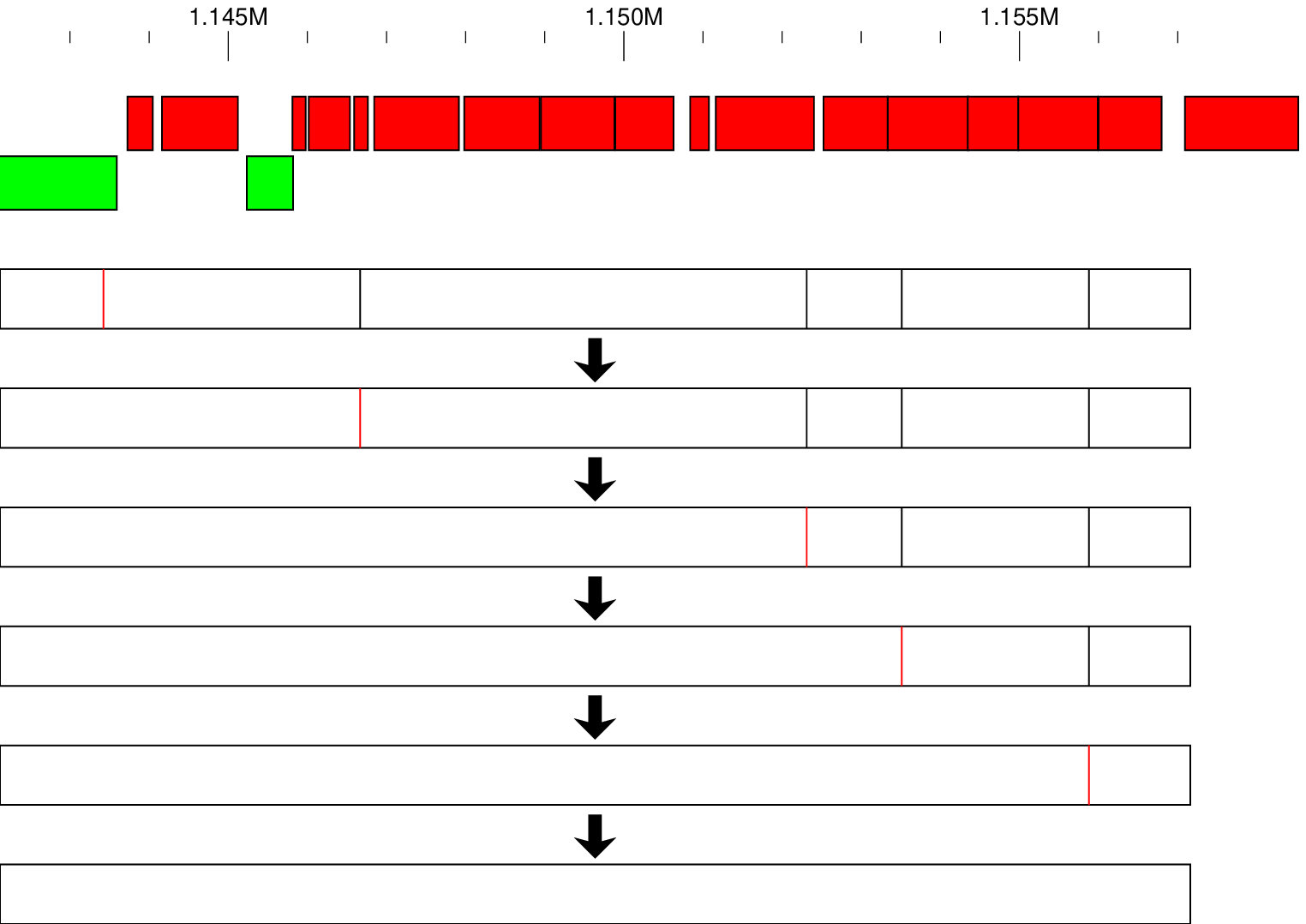}
\caption{A cascade event occuring between $M = 1846$ and $M = 1841$
domain walls remaining in the bottom-up segmentation history of
\emph{E. coli} K-12 MG1655 ($N = 4639675$ bp), starting from the $(K =
0, n = 1000)$ initial segmentation containing $M = 2791$ domain walls.
Six segments in the $(1142115, 1157158)$ region of the genome are
shown.  The first domain wall to be removed in this cascade event lies
close to the boundary between the gene \emph{rne}, believed to be
RNase E, on the negative strand (green), and the gene \emph{yceQ},
coding for a hypothetical protein, on the positive strand (red).  The
second domain wall to be removed in the cascade is in the middle of
the gene \emph{rpmF} on the positive strand, the third is close to the
boundary between \emph{fabF} and \emph{pabC}, the fourth is close to
the boundary between \emph{pabC} and \emph{yceG}, and the last is
close to the boundary between \emph{holB} and \emph{ycfH}.  At each
stage of the bottom-up segmentation history, the domain wall removed
is highlighted in red.}
\label{figure:NC000913topsegKL1846}
\end{figure}

Looking more closely at the segment statistics, we realized that a
tunneling event involves a short segment flanked by two long segments
which are statistically similar to one another, but different from the
short segment.  This statistical dissimilarity between the short
segment and its long flanking segments is reflected in the moderate
strengths $\Delta_L$ and $\Delta_R$ of the left and right domain walls
of the short segment.  Let us say the right domain wall is slightly
weaker than the left domain wall, i.e. $\Delta_R \lesssim \Delta_L$.
As the bottom-up segmentation history progresses, there will reach a
stage where we remove the right domain wall.  When this happens, the
short segment will be assimilated by its right flanking segment.
Because the right flanking segment is long, absorbing the short
segment represents only a small perturbation in its segment
statistics.  The longer right segment that results is still
statistically similar to the left segment.  Therefore, when we
recompute the strength $\Delta_L$ of the remaining domain wall, we
find that it is now smaller than the strength $\Delta_R$ of the domain
wall that was just removed.  This remaining domain wall therefore
becomes the next to be removed in the bottom-up segmentation history,
afterwhich the next domain wall to be removed occurs somewhere else in
the sequence, and has strength slightly larger than $\Delta_R$.  The
signature of a tunneling event is therefore a sharp dip in the
bottom-up segmentation history.  Biologically, a short segment with a
tunneling event signature is likely to represent an insertion sometime
in the evolutionary past of the organism.  A tunneling event in the
$(K = 0, n = 1000)$ bottom-up segmentation history is shown in Fig.
\ref{figure:NC000913topsegKL1586}.  In contrast, a cascade event
involves a cluster of short segments of varying statistics flanked by
two long segments that are statistically similar.  The domain walls
separating the short segments from each other and from the long
flanking segments are then removed in succession.  This sequential
removal of domain walls gives rise to an extended dip in the bottom-up
segmentation history, with a complex internal structure that depends
on the actual distribution of short segments.  Biologically, a cluster
of short segments participating in a cascade event points to a
possible recombination hotspot on the genome of the organism.  A
cascade event in the $(K = 0, n = 1000)$ bottom-up segmentation
history is shown in Fig.  \ref{figure:NC000913topsegKL1846}. 

Clearly, by removing more and more domain walls, we construct a proper
hierarchy of segmentations containing fewer and fewer domain walls,
which agrees intuitively with our notion of what coarse graining is
about.  We also expected to obtain a unique coarse-grained
segmentation, containing only domain walls stronger than $\Delta^*$,
by removing all domain walls weaker than $\Delta^*$.  It turned out
the picture that emerge from this coarse graining procedure is more
complicated, based on which we identified three main problems.  First,
let us start with a segmentation containing domain walls weaker than
$\Delta^*$, and decide to remove these domain walls in a single step.
Recomputing the strengths of the remaining domain walls, we would find
that some of these will be weaker than $\Delta^*$, and so cannot claim
to have found the desired coarse-grained segmentation.  Naturally, we
iterate the process, removing all domain walls weaker than $\Delta^*$,
and recomputing the strengths of the remaining domain walls, until all
remaining domain walls are stronger than $\Delta^*$.  Next, we try
removing domain walls weaker than $\Delta^*$ one at a time, starting
from the weakest, and recompute domain wall strengths after every
removal.  The strengths of a few of the remaining domain walls will
change each time the weakest domain wall is removed, sometimes
becoming stronger, and sometimes becoming weaker, but we continue
removing the weakest domain wall until all remaining domain walls are
stronger than $\Delta^*$.  Comparing the segmentations obtained using
the two coarse-graining procedures, we will find that they can be very
different.  This difficulty occurs for all averaging problems, so we
are not overly concerned, but argue instead that removing the weakest
domain wall each time is like a renormalization-group procedure, and
should therefore be more reliable than removing many weak domain walls
all at once.  

Once we accept this decremental procedure for coarse graining, we
arrive at the second problem.  Suppose we do not stop coarse graining
after arriving at the first segmentation with all domain walls
stronger than $\Delta^*$, but switch strategy to target and removing
segments associated with tunneling and cascade events.  The
segmentations obtained after all domain walls associated with such
segments will contain only domain walls stronger than $\Delta^*$, but
the segmentations in the intermediate steps will contain domain walls
weaker than $\Delta^*$.  If we keep coarse graining until no tunneling or
cascade events weaken domain walls below $\Delta^*$, we would end up
with a series of coarse-grained segmentations containing different
number of domain walls.  These segmentations do not have the same
minimum domain wall strengths, but are related to each other through
stages in which some domain walls are weaker than $\Delta^*$.  We
worry about this series of segmentations when there exist domain walls
with equal or nearly equal strengths.  If at any stage of the coarse
graining, these domain walls become the weakest overall, and we stick
to removing one domain wall at a time, we can remove any one of these
equally weak domain walls.  If we track the different bottom-up
segmentation histories associated with each choice, we will find that
the coarse-grained segmentations for which all domain walls first
become stronger than $\Delta^*$ can be very different.  However, if we
coarse grain further by targetting tunneling and cascading segments,
we would end up with the same coarse-grained segmentation for which no
domain walls ever become weaker than $\Delta^*$.  Another way to think
of this coarsest segmentation is that it is the one for which no
domain wall stronger than $\Delta^*$ can be added without first adding
a domain wall weaker than $\Delta^*$.

Third, we know from the bottom-up segmentation history that short
segments participating in tunneling events can be absorbed into their
long flanking segments without appreciably changing the strengths of
the latter's other domain walls.  Clearly, absorbing statistically
very distinct short segments increases the heterogenuity of the
coarse-grained segment.  This is something we have to accept in coarse
graining, but ultimately, what we really want at each stage of the
coarse graining is for segments to be no more heterogeneous than some
prescribed segment variance.  Unfortunately, the segment variances are
not related to the domain wall strengths in a simple fashion, and even
if we know how to compute these segment variances, there is no
guarantee that a coarse graining scheme based on these will be less
problematic.  The bottomline is, all these problems arise because
domain wall strengths change wildly as segments are agglomerated in
the coarse graining process, due again to the context sensitivity of
the Jensen-Shannon divergence (or any other entropic measure, for that
matter).

\section{Mean-Field Analyses of Segmentation Schemes}
\label{section:meanfieldanalysis}

From our segmentation and coarse graining analyses of real genomes in
Sec.~\ref{section:contextsensitivityprobleminrealgenomes}, we realized
that these cannot be thought of as consisting of long segments that
are strongly dissimilar to its neighboring long segments, within which
we find short segments that are weakly dissimilar to its neighboring
short segments.  In fact, the results suggest that there are short
segments that are strongly dissimilar to its neighboring long
segments, which are frequently only weakly dissimilar to its
neighboring long segments.  This mosaic and non-hierarchical structure
of segments is the root of the context sensitivity problem, which we
will seek to better understand in this section.

\begin{figure}[htbp]
\centering
\includegraphics[width=.7\linewidth]{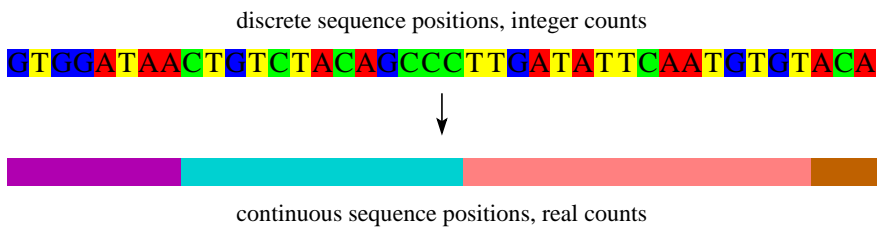}
\caption{Going from a discrete description to a continuum description
of a nucleotide sequence.}
\label{figure:meanfieldpicture}
\end{figure}

To do this, we go first to a continuum description of discrete genomic
sequences, as shown in Fig.~\ref{figure:meanfieldpicture}, where we
allow the sequence positions and the various $K$-mer frequencies to
vary continuously.  To eliminate spatial inhomogenuities in the
statistics of the interval $[i, j > i)$, which we want to model as a
statistically stationary segment in the \emph{mean-field limit}, we
distribute its $K$-mer statistics uniformly along the segment.  More
precisely, if $f_{\bt s}^{[i, j)}$ is the number of times the
$(K+1)$-mer $\alpha_{t_K}\cdots\alpha_{t_1}\alpha_s$, which we also
refer to as the \emph{transition} $\bt \to s$, appears in $[i, j)$, we
define the mean-field count $f_{\bt s}^{[i', j')}$ of the transition
$\mathbf{t} \to s$ within the subinterval $[i', j' > i') \subseteq [i,
j)$ to be
\begin{equation}
f_{\mathbf{t}s}^{[i', j')} \equiv
\frac{j' - i'}{j - i}\,
f_{\mathbf{t}s}^{[i, j)}.
\end{equation}
Within this mean-field picture, we discuss in
Sec.~\ref{subsection:meanfieldwindowedspectrum} how the paired
sliding-window scheme's ability to detect domain walls depends on the
size $n$ of the pair of sliding windows.  We show, in contrast to the
positions and strengths being determined exactly by this segmentation
scheme for domain walls between segments both longer than $n$, that
domain walls between segments, one or both of which are shorter than
$n$, are weakened and shifted in the mean-field limit.  Following
this, we show in Sec.~\ref{subsection:meanfieldwindowlessspectrum}
that the strengths of the domain walls obtained from the recursive
segmentation scheme are context sensitive, and approach the exact
strengths only as we approach the terminal segmentation.  We explain
why optimization is desirable at every step of the recursive
segmentation, before going on to explain why repetitive sequences are
the worst kind of sequences to segment in
Sec.~\ref{subsection:oscillatorysequence}.  In this section, we
present numerical examples for $K = 0$ Markov chains, but all
qualitative conclusions are valid for Markov chains of order $K > 0$.

\subsection{Paired Sliding Windows Segmentation Scheme}
\label{subsection:meanfieldwindowedspectrum}

\begin{figure}[htbp]
\centering
\includegraphics{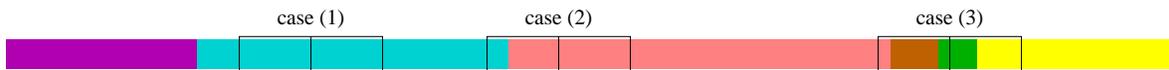}
\caption{The three possible situations that we encounter when we slide
a symmetric pair of windows across a sequence composed of many
mean-field segments: (1) both windows lie entirely within a single
mean-field segment; (2) the two windows straddle two mean-field
segments; and (3) the two windows straddle multiple mean-field
segments.}
\label{figure:windowedcases}
\end{figure}

For a pair of windows of length $n$ sliding across a mean-field
sequence, there are three possibilities (see
Fig.~\ref{figure:windowedcases}):
\begin{enumerate}

\item both windows lie entirely within a single mean-field segment;

\item the two windows straddle two mean-field segments, i.e. a single
domain wall within one of the windows;

\item the two windows straddle multiple mean-field segments.

\end{enumerate}
The first situation is trivial, as the left and right windowed counts
are identical,
\begin{equation}
f_{\mathbf{t}s}^L = f_{\mathbf{t}s}^R = \frac{n}{N_{\text{seg}}}\,
f_{\mathbf{t}s}^{\text{seg}},
\end{equation}
$N_{\text{seg}}$ being the length of the mean-field segment, and
$f_{\mathbf{t}s}^{\text{seg}}$ being the transition counts within the
mean-field segment.  The Jensen-Shannon divergence, or the square
deviation between the two windows therefore vanishes identically.  The
second situation, which is what the paired sliding windows
segmentation scheme is designed to handle, is analyzed in
App.~\ref{subsection:meanfieldlineshape}.  Based on that analysis, we
showed that the position and strength of the domain wall between the
two mean-field segments can be determined exactly.  We also derived
the mean-field lineshape for match filtering.

\begin{figure}[htbp]
\centering
\includegraphics[scale=0.45,clip=true]{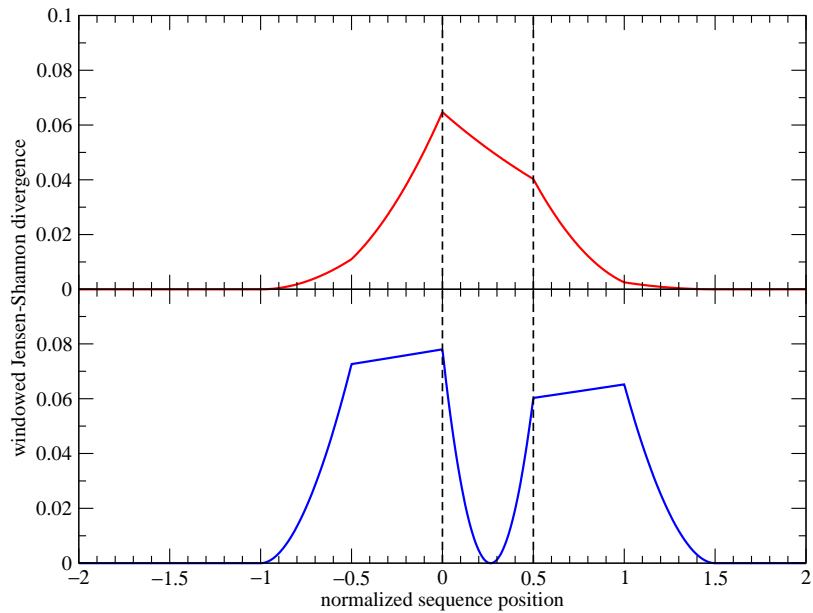}
\caption{The Jensen-Shannon divergence $\Delta(z)$ (solid curves) of a
pair of sliding windows of length $n = 1$ as it slides across the
binary mean-field segments (left to right) $a$, $b$, and $c$, with
lengths $N_a > 1$, $N_b < 1$, and $N_c > 1$ respectively.  On the
above plots, the left and right ends of segment $b$ are highlighted by
the dashed vertical lines at the normalized sequence positions $z = 0$
and $z = 0.5$ respectively.  For the top plot, the probabilities
associated with the mean-field segments are $P_a(0) = 1 - P_a(1) =
0.30$, $P_b(0) = 1 - P_b(1) = 0.50$, and $P_c(0) = 1 - P_c(1) = 0.60$.
For the bottom plot, the probabilities associated with the mean-field
segments are $P_a(0) = 1 - P_a(1) = 0.20$, $P_b(0) = 1 - P_b(1) =
0.70$, and $P_c(0) = 1 - P_c(1) = 0.22$.}
\label{figure:wJSshortsegment}
\end{figure}

In this subsection, our interest is in understanding how the paired
sliding windows segmentation scheme behaves in the third situation.
Clearly, the precise structure of the mean-field divergence spectrum
will depend on the local context the pair of windows is sliding
across, so we look at an important special case: that of a pair of
length-$n$ windows sliding across a segment shorter than $n$.  In Fig.
\ref{figure:wJSshortsegment}, we show two lineshapes which are
expected to be generic, for (i) the long segments flanking the short
segment are themselves statistically dissimilar (top plot); and (ii)
the long segments flanking the short segment are themselves
statistically similar (bottom plot).  In case (i), the mean-field
lineshape obtained as the pair of windows slides across the short
segment consists of a single peak at one of its ends.  This peak is
broader than that of a simple domain wall by the width of the short
segment, and therefore, if we perform match filtering using the
quadratic mean-field lineshape in Eq.  \eqref{equation:meanfieldJS},
the center of the match-filtered peak would occur not at either ends
of the short segment, but somewhere in the interior.  

In case (ii), the mean-field lineshape obtained as the pair of windows
slides across the short segment consists of a pair of peaks, both of
which are narrower than the mean-field lineshape of a single domain
wall. After we perform match filtering, the center of the
match-filtered left peak would be left of the true left domain wall,
while the center of the match-filtered right peak would be right of
the true right domain wall.  Case (ii) is of special interest to us,
as it is the context that give rise to tunneling events in the
bottom-up segmentation history.  Both contexts give rise to shifts in
the domain wall positions, as well as to changes in the strengths of
the unresolved domain walls, and thus may be able to explain some of
the observations made in Sec.~\ref{subsection:pairedslidingwindows}.
In case (i), the domain wall strength can increase or decrease,
depending on how different the two long flanking segments are compared
to the short segment.  In case (ii), the domain wall strengths always
decrease.

\subsection{Optimized Recursive Jensen-Shannon Segmentation Scheme}
\label{subsection:meanfieldwindowlessspectrum}

\begin{figure}[htbp]
\centering
\includegraphics[scale=0.45,clip=true]{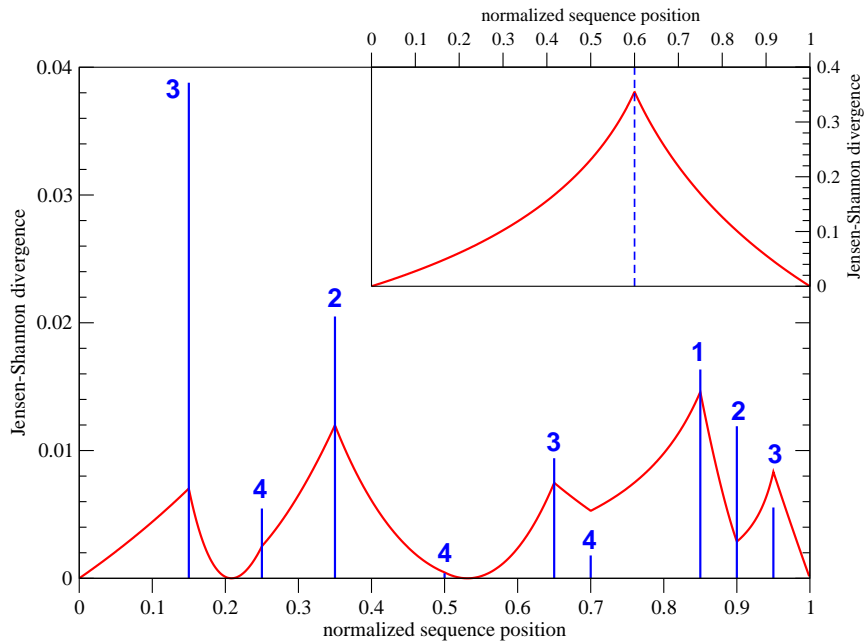}
\caption{The Jensen-Shannon divergence $\Delta(z)$ (red solid curve)
as a function of the normalized cursor position $z$ within an
artificial binary sequence composed of ten mean-field segments,
characterized by the probabilities (left to right) $\mathbf{P}(0) =
(0.55, 0.05, 0.20, 0.60, 0.65, 0.30, 0.45, 0.05, 0.45, 0.15)$.  The
blue bars indicate the true strengths of each of the nine domain
walls, at $z_1 = 0.15$, $z_2 = 0.25$, $z_3 = 0.35$, $z_4 = 0.50$, $z_5
= 0.65$, $z_6 = 0.70$, $z_7 = 0.85$, $z_8 = 0.90$, and $z_9 = 0.95$,
while the number at each domain wall indicate which recursion step it
is discovered.  (Inset) The Jensen-Shannon divergence $\Delta(z)$
(red solid curve) as a function of the normalized cursor position $z$
within an artificial binary sequence composed of two mean-field
segments, characterized by the probabilities $P_L(0) = 0.10$ and
$P_R(0) = 0.90$.  The domain wall at $z = 0.60$ is indicated by the
blue dashed vertical line.}
\label{figure:JS10}
\end{figure}

To understand how the optimized recursive Jensen-Shannon segmentation
is sensitive to global context, let us first understand what happens
when the segments discovered recursively are not optimized, and then
consider the effects of segmentation optimization.  In
Fig.~\ref{figure:JS10}, we show the Jensen-Shannon
divergence spectrum for a sequence consisting of ten mean-field
segments.  As we can see, the mean-field Jensen-Shannon divergence is
everywhere convex, except at the domain walls.  These are associated
with peaks or kinks in the divergence spectrum, depending on the
global context within the sequence.  Under special distributions of
the segment statistics, domain walls may even have vanishing
divergences.

All nine domain walls in the ten-segment sequence are recovered if we
allow the recursive Jensen-Shannon segmentation without segmentation
optimization to go to completion.  However, as shown in
Fig.~\ref{figure:JS10}, these domain walls are not discovered in the
order of their true strengths (heights of the blue bars), given by
the Jensen-Shannon divergence between the pairs of segments they
separate.  In fact, just like in the coarse graining procedure
described in Sec.~\ref{subsection:bottomupsegmentationhistory}, the
Jensen-Shannon divergence at each domain wall changes as the recursion
proceeds, as the context it is found in gets refined.  For this
ten-segment sequence, the recursive segmentation scheme's sensitivity
to global context results in the third strongest domain wall being
discovered in the first recursion step, the second and fourth
strongest domain walls being discovered in the second recursion step,
and the strongest domain wall being discovered only in the third
recursion step.

\begin{figure}[htbp]
\centering
\includegraphics[scale=0.45,clip=true]{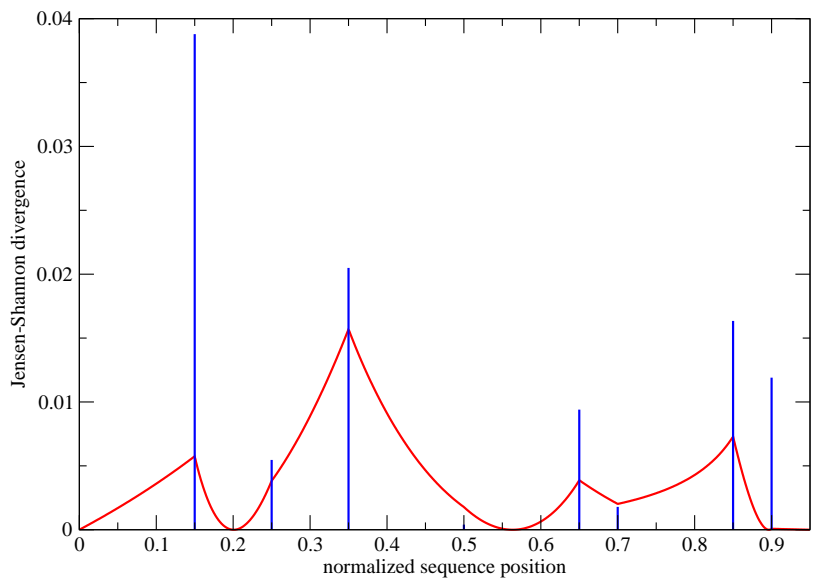}
\caption{The windowless Jensen-Shannon divergence spectrum $\Delta(z)$
(red solid curve) of the nine-segment binary sequence, after losing
the short segment at its right end.  The blue bars indicate the
strength of each of the nine domain walls.}
\label{figure:JS10c9}
\end{figure}

To see the extent to which optimization ameliorate the global context
sensitivity of the recursive segmentation scheme, let us imagine the
ten-segment sequence to be part of a longer sequence being recursively
segmented.  Let us further suppose that under segmentation
optimization, the segment $(0.95, 1.00)$ gets incorporated by the
sequence to the right of $(0.00, 1.00)$.  With this, we now examine in
detail a nine-segment sequence $(0.00, 0.95)$, whose mean-field
divergence spectrum is shown in Fig.~\ref{figure:JS10c9}, instead of
the original ten-segment sequence $(0.00, 1.00)$.  From
Fig.~\ref{figure:JS10c9}, we find the divergence maximum of the
nine-segment sequence is at $z_3 = 0.35$, the second strongest of the
nine domain walls, instead of the third strongest domain wall at $z_7
= 0.85$ for the ten-segment sequence.  In proportion to the length of
the ten-segment sequence, this shift from the third strongest domain
wall to the second strongest domain wall is huge, by about half the
length of the sequence, when the change in context involves a loss of
only 5\% of the total length.  In
Sec.~\ref{subsection:recursivegenome}, we saw instances of such large
shifts in optimized domain wall positions when we recursively add one
new domain wall each time to a real genome.

In this example of the ten-segment sequence, we saw that segmentation
optimization has the potential to move an existing domain wall, from a
weaker (the third strongest overall), to a stronger (the second
strongest overall, and if the global context is different, perhaps even
to the strongest overall) position.  However, the nature of the
context sensitivity problem is such that no guarantee can be offered
on the segmentation optimization algorithm always moving a domain wall
from a weaker to a stronger position.  Nevertheless, segmentation
optimization frequently does move a domain wall from a weaker position
to a stronger position, and it always make successive segments as
statistically distinct from each other as possible.  This is good
enough a reason to justify the use of segmentation optimization.

\subsection{Repetitive Sequences}
\label{subsection:oscillatorysequence}

In this last subsection of Sec.~\ref{section:meanfieldanalysis}, let
us look at repetitive sequences, for which the context sensitivity
problem is the most severe.  Such sequences, which are composed of
periodically repeating motifs, are of biological interest because they
arise from a variety of recombination processes, and are fairly common
in real genomic sequences.  In general, a motif $a_1a_2\cdots a_r$
that is repeated in a repetitive sequence can consists of $r$
statistically distinct subunits, but for simplicity, let us look only
at $ab$-repeats, and highlight statistical signatures common to all
repetitive sequences.

\begin{figure}[htbp]
\centering
\includegraphics[scale=0.45,clip=true]{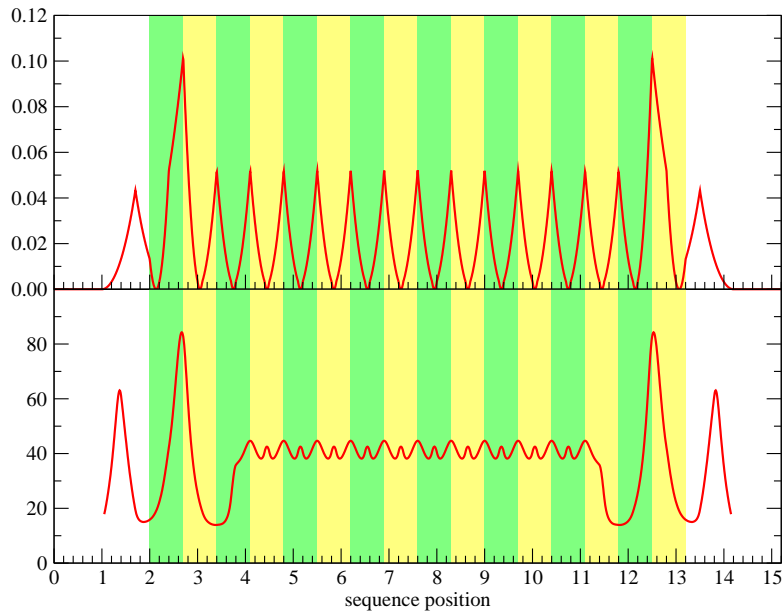}
\caption{The Jensen-Shannon divergence spectrum (top, red solid curve)
before, and (bottom, red solid curve) after match filtering and
quality enhancement, for a pair of windows of size $n = 1$ sliding
across a repetitive binary $K = 0$ sequence $cababababababababc$,
where the subunits $a$ (light green) and $b$ (light yellow) both have
lengths $n_a = n_b = 0.7$, and are characterized by the probabilities
$P_a(0) = 1 - P_a(1) = 0.1$ and $P_b(0) = 1 - P_b(1) = 0.9$.  The
terminal $c$ segments (white), assumed to have lengths much larger
than $n = 1$, are characterized by the probability $P_c(0) = 1 -
P_c(1) = 0.5$.}
\label{figure:wabx8mf}
\end{figure}

When we segment the repetitive sequence $abababababababab$ using the
paired sliding windows segmentation scheme with window size $n$, we
obtained the mean-field Jensen-Shannon divergence spectrum shown in
the top plot of Fig.~\ref{figure:wabx8mf}.  In this figure, sequence
positions are normalized such that $n = 1$, while the lengths of the
repeating segments $a$ and $b$ are chosen to be both less than the
window size, i.e.~$n_a = n_b = 0.7 < n$.  To understand contextual
effects at the ends of the repetitive sequence, we include the
terminal segments $c$ in our analysis.  These terminal $c$ segments
are assumed to have lengths $n_c \gg n$, and statistics intermediate
between those of $a$ and $b$.  As we can see from the top plot of
Fig.~\ref{figure:wabx8mf}, all domain walls between $a$ and $b$
segments ($ab$ \emph{domain walls}) correspond to peaks in the
mean-field divergence spectrum.  The two $ab$ domain walls near the
ends of the repetitive sequence are the strongest, while the rest have
the same diminished strength (compared to the Jensen-Shannon
divergence between the $a$ and $b$ segments).  From the top plot of
Fig.~\ref{figure:wabx8mf}, we also see that no peaks are associated
with the $ca$ and $bc$ domain walls.  Instead, we find a spurious peak
left of the $ca$ domain wall, and another spurious peak right of the
$bc$ domain wall.

As discussed in
App.~\ref{section:pairedslidingwindowssegmentationscheme}, the
mean-field lineshape of a simple domain wall is very nearly piecewise
quadratic, with a total width of $2n$.  This observation is extremely
helpful when we deal with real divergence spectra, where statistical
fluctuations produce spurious peaks with various shapes and widths.
By insisting that only peaks that are (i) approximately piecewise
quadratic, with (ii) widths close to $2n$, are statistically
significant, we can determine a smaller, and more reliable set of
domain walls through match filtering.  In the top plot of
Fig.~\ref{figure:wabx8mf}, all our peaks have widths smaller than
$2n$.  In the mean-field limit, these are certainly not spurious, but
if we imagine putting statistical fluctuations back into the
divergence spectrum, and suppose we did not know beforehand that there
are segments shorter than $n$ in this sequence, it would be reasonable
to accept by fiat whatever picture emerging from the match filtering
procedure.  For $cababababababababc$, the match-filtered, quality
enhanced divergence spectrum is shown as the bottom plot of
Fig.~\ref{figure:wabx8mf}, where we find the two spurious peaks
shifted deeper into the $c$ segments by the match filtering procedure.
In this plot, the two strong $ab$ domain walls near the ends of the
repetitive sequence continue to stand out, but the rest of the $ab$
domain walls are now washed out by match filtering.  If we put
statistical noise back into the picture, the fine structures marking
these remaining $ab$ domain walls will disappear, and we end up with a
featureless plateau in the interior of the repetitive sequence.  We
might then be misled into thinking that this $cababababababababc$
sequence consists of only five segments $ca'c'b'c$, where $a'$ is $a$
contaminated by a small piece of $c$, $b'$ is $b$ contaminated by a
small piece of $c$, and $c'$, which lies between the two strong $ab$
domain walls, will be mistaken for a segment with $K = 0$ statistics
similar to $c$, even though it is not statistically stationary.  

\begin{figure}[htbp]
\centering
\includegraphics[scale=0.45,clip=true]{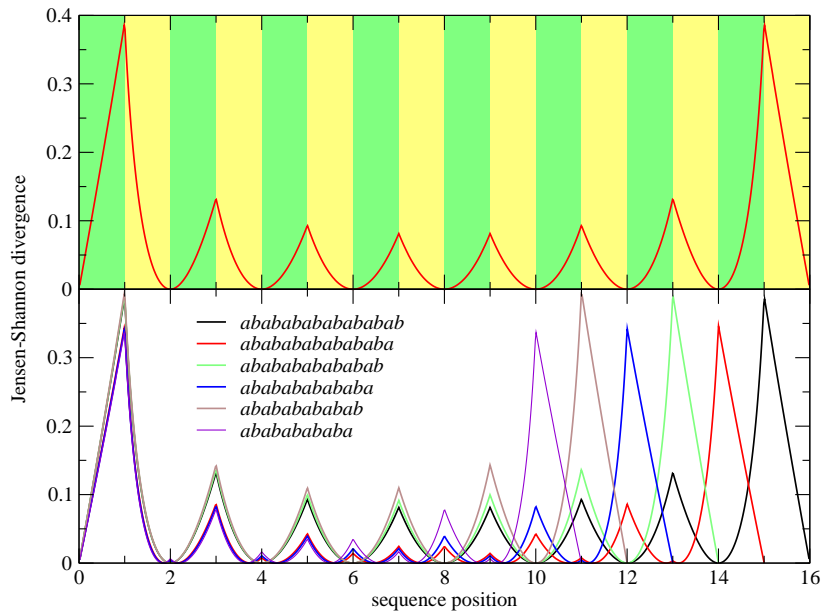}
\caption{(Top) The top-level Jensen-Shannon divergence spectrum (red
solid curve) obtained in the recursive segmentation of a repetitive
binary sequence consisting of subunits $a$ (light green, $P_a(0) = 1 -
P_a(1) = 0.1$) and $b$ (light yellow, $P_b(0) = 1 - P_b(1) = 0.9$)
repeated eight times.  (Bottom) The Jensen-Shannon divergence spectra
obtained when $abababababababab$ is recursively segmented from the
right end.}
\label{figure:abx8rJS}
\end{figure}

Next, let us analyze the recursive Jensen-Shannon segmentation of
$abababababababab$, where we cut the repetitive sequence first into
two segments, then each of these into two subsegments, and so on and
so forth, until all the segments are discovered.  In the top plot of
Fig.~\ref{figure:abx8rJS}, we show the top-level Jensen-Shannon
divergence spectrum, based on which we will cut $abababababababab$
into two segments.  In this figure, we find 
\begin{enumerate}

\item a series of $k$ peaks of unequal strengths, with stronger peaks
near the ends, and weaker peaks in the middle of the repetitive
sequence;

\item $k - 1$ domain walls having vanishing divergences;

\item the ratio of strengths of the strongest peak to the weakest peak
is roughly $k/2$,

\end{enumerate}
where $k$ is the number of repeated motifs.  These statistical
signatures are shared by all repetitive sequences, with the detail
distribution and statistical characteristics of the subunits within
the repeated motif affecting only the shape and strength of the peaks.
Here we see extreme context sensitivity reflected in the fact that
domain walls with the same true strength can have very different, and
even vanishing, strengths when the segment structure of the sequence
is examined recursively.

\begin{figure}[htbp]
\centering
\includegraphics[scale=0.45,clip=true]{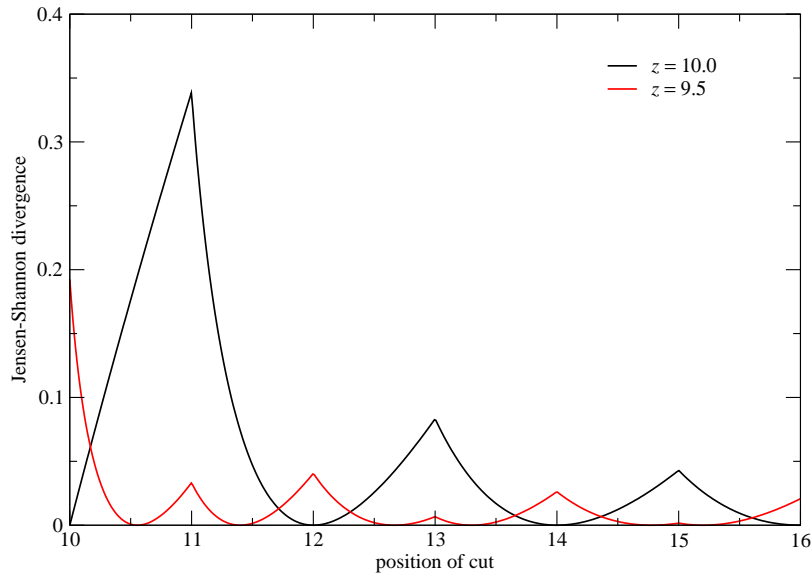}
\caption{The windowless Jensen-Shannon divergences at $z = 10.0$ (at a
domain wall) and $z = 9.5$ (away from a domain wall) of the repetitive
binary sequence $abababababababab$, with $P_a(0) = 0.1$ and $P_b(0) =
0.9$, as functions of the cut $10 \leq z \leq 16$.}
\label{figure:abx8osc}
\end{figure}

From the bottom plot of Fig.~\ref{figure:abx8rJS}, we find that one or
both of the peaks near the ends of the repetitive sequence are always
the strongest, as recursion progresses.  This is true when the
repetitive sequence consists of repeating motifs with more complex
internal structure, and also true when we attach terminal segments to
the repetitive sequence.  Therefore, successive cuts are always made
at one end or the other of the repetitive sequence.  For $ab$-repeats,
the peaks near both ends are equally strong in the mean-field limit,
so we can choose to always cut at the right end of $abababababababab$,
as shown in the bottom plot of Fig.~\ref{figure:abx8rJS}.  As the
repetitive sequence loses its rightmost segment at every step, and the
global context alternates between being dominated by $a$ segments to
being dominated by $b$ segments, we find oscillations in the strengths
of the remaining domain walls.  This oscillation, which is a generic
behaviour of all repetitive sequences under recursive segmentation,
can be seen more clearly for the $ab$-repetitive sequence in Figure
\ref{figure:abx8osc}, where instead of cutting off one segment at a
time, we move the cut continuously inwards from the right end.

\section{Summary and Discussions}
\label{section:conclusions}

In this paper, we defined the \emph{context sensitivity problem}, in
which the \emph{same} group of statistically stationary segments are
segmented \emph{differently} by the \emph{same} segmentation scheme,
when it is encapsulated within \emph{different larger contexts} of
segments.  We then described in
Sec.~\ref{section:contextsensitivityprobleminrealgenomes} the various
manifestions of context sensitivity when real bacterial genomes are
segmented using the paired sliding windows and optimized recursive
Jensen-Shannon segmentation schemes, which are sensitive to local and
global contexts respectively.  For the single-pass paired sliding
windows segmentation scheme, we found that the positions and relative
strengths of domain walls can change dramatically when we change the
window size, and hence the local contexts examined.  For the optimized
recursive segmentation scheme, we found that there can be large shifts
in the optimized domain wall positions as recursion progresses, due to
the change in global context when we go from examining a sequence to
examining its subsequence, and \emph{vice versa}.

In Sec.~\ref{section:contextsensitivityprobleminrealgenomes}, we also
looked into the issue of coarse graining the segmental description of
a bacterial genome.  We argued that coarse graining by using larger
window sizes, or stopping recursive segmentation earlier can be
biologically misleading, because of the context sensitivity problem,
and explored an alternative coarse graining procedure which involves
removing the weakest domain walls and agglomerating the segments they
separate.  This coarse graining procedure was found to be fraught with
difficulties, arising again from the context sensitivity of domain
wall strengths.  Ultimately, the goal of coarse graining is to reduce
the complexity of the segmented models of real genomes.  This can be
achieved by reducing the number of segments, or by reducing the number
of segment \emph{types} or \emph{classes} (see, for example, the work
by Azad \emph{et al}.  \cite{Azad2002PhysicalReviewE66a031913}).  We
realized in this paper that the former is unattainable, and proposed
to accomplish the latter through statistical clustering of the
segments.  Based on what we understand about the context sensitivity
problem, we realized that it would be necessary to segment a given
genomic sequence as far as possible, to the point before genes are cut
into multiple segments (unless they are known to contain multiple
domains).  We are in the process of writing the results of our
investigations into this manner of coarse graining, in which no domain
walls are removed, but statistically similar segments are clustered
into a small number of segment classes.

In Sec.~\ref{section:meanfieldanalysis}, we analyzed the paired
sliding windows and optimized recursive segmentation schemes within a
mean-field picture.  For the former, we explained how the presence of
segments shorter than the window size lead to shifts in the positions,
and changes in the strengths of domain walls.  For the latter, we
illustrate the context dependence of the domain walls strengths, how
this leads to large shifts in the optimized domain wall positions, and
also to the domain walls being discovered out of order by their true
strengths.  We showed that all domain walls in a sequence will be
recovered in the mean-field limit, if we allow the recursive
segmentation to go to completion, but realized that for real sequences
subject to statistical fluctuations, there is a danger of stopping the
the recursion too early.  When this happens, we will generically pick
up weak domain walls, but miss stronger ones --- a problem that can be
partly alleviated through segmentation optimization, in which domain
walls are moved from weaker to stronger positions.  We devoted one
subsection to explain why the context sensitivity problem is
especially severe in repetitive sequences.

Finally, let us say that while we have examined only two entropic
segmentation schemes in detail, we believe the context sensitivity
problem plagues all segmentation schemes.  The manifestations of the
context sensitivity problem will of course be different for different
segmentation schemes, but will involve (i) getting the domain wall
positions wrong; (ii) getting the domain wall strengths wrong; or
(iii) missing strong domain walls.  A proper analysis of the context
sensitivity of the various segmentation schemes is beyond the scope of
this paper, but let us offer some thoughts on segmentation schemes based
on based on hidden Markov models (HMMs), which are very popular in the
bioinformatics literature.  In HMM segmentation, model parameters are
typically estimated using the Baum-Welch algorithm, which first
computes the forward and backward probabilities of each hidden state,
use these to estimate the transition frequencies, which are used to
update the model parameters.  Computation of forward and backward
probabilities are sensitive to local context, in that the hidden
states assigned to a given collection of segments will be different,
if the sequences immediately flanking the segments are different.
Updating of model parameters, on the other hand, is sensitive to
global context, because very different arrangement of segments and
segment classes can give rise to the same summary of transition
frequencies.  The signatures of this dual local-global context
sensitivity is buried within the sequence of posterior probabilities
obtained from iterations of the Baum-Welch algorithm.  Ultimately, the
context sensitivity problem is a very special case of the problem of
mixed data, which is an active area of statistical research.  We hope
that through the results presented in this paper, the bioinformatics
community will come to better recognize the nuances sequence context
poses to its proper segmentation.

\begin{appendix}

\subsection{Generalized Jensen-Shannon Divergences}
\label{section:generalizedJensenShannondivergences}

In Ref.~\citeonline{Cheong2007IRJSSS} we explained that dinucleotide
correlations and codon biases in biological sequences
\cite{Grantham1981NucleicAcidsResearch9pR43,
Shepherd1981ProcNatlAcadSciUSA78p1596,
Staden1982NucleicAcidsResearch10p141,
Fickett1982NucleicAcidsResearch10p5303, Herzel1995PhysicaA216p518} are
better modeled by Markov chains of order $K > 0$ over the quaternary
alphabet $\mathcal{S} = \rm\{A, C, G, T\}$
\cite{Thakur2007PhysicalReviewE75a011915}, rather than Bernoulli
chains over $\mathcal{S}$
\cite{BernaolaGalvan1996PhysicalReviewE53p5181,
RomanRoldan1998PhysicalReviewLetters80p1344}, or Bernoulli chains over
the extended alphabet $\mathcal{S}^K$
\cite{BernaolaGalvan2000PhysicalReviewLetters85p1342,
Nicorici2003FINSIG03,
Nicorici2004EURASIPJournalonAppliedSignalProcessing1p81}.   In the
sequence segmentation problem, our task is to decide whether there is
a domain wall at sequence position $i$ within a given sequence
$\bx = x_1 x_2 \cdots x_{i-1} x_i x_{i+1} \cdots x_N$, where $x_j \in
\mathcal{S}, 1 \geq j \geq N$.  The simplest model selection scheme
that would address this problem would involve the comparison of the
one-segment sequence likelihood $P_1$, whereby the sequence $\bx$ is
treated as generated by a single Markov process, against the
two-segment sequence likelihood $P_2$, whereby the subsequences $\bx_L
= x_1 x_2 \cdots x_{i-1}$ and $\bx_R = x_i x_{i+1} \cdots x_N$ are
treated as generated by two different Markov processes.

To model $\bx$, $\bx_L$, and $\bx_R$ as Markov chains of order $K$, we
determine the order-$K$ \emph{transition counts} $f_{\bt s}$, $f_{\bt
s}^L$, $f_{\bt s}^R$, subject to the normalizations
\begin{equation}
f_{\bt s} = f_{\bt s}^L + f_{\bt s}^R, \quad
\sum_{\bt\in\mathcal{S}^K}\sum_{s=1}^S f_{\bt s} = N.
\end{equation}
Here $S = 4$ is the size of the quaternary alphabet $\mathcal{S}$, and
$\bt$ is a shorthand notation for the $K$-tuple of indices $(t_1, t_2,
\dots, t_K), 1 \leq t_k \leq S$.  The transition counts $f_{\bt s}$,
$f_{\bt s}^L$, and $f_{\bt s}^R$ are the number of times the
$(K+1)$-mer $\alpha_{t_K} \cdots \alpha_{t_1} \alpha_s$ appear in the
sequences $\bx$, $\bx_L$, and $\bx_R$ respectively.  The sequences
$\bx$, $\bx_L$, and $\bx_R$ are then assumed to be generated by the
Markov processes with \emph{maximum-likelihood transition
probabilities}
\begin{equation}
\hat{p}_{\bt s} = \frac{f_{\bt s}}{\sum_{s'=1}^S f_{\bt s'}}, \quad
\hat{p}_{\bt s}^L = \frac{f_{\bt s}^L}{\sum_{s'=1}^S f_{\bt s'}^L}, \quad
\hat{p}_{\bt s}^R = \frac{f_{\bt s}^R}{\sum_{s'=1}^S f_{\bt s'}^R},
\end{equation}
respectively.

Within these maximum-likelihood Markov-chain models, the one- and
two-segment sequence likelihoods are given by
\begin{equation}
\begin{aligned}
P_1 &= \prod_{\bt\in\mathcal{S}^K}\prod_{s=1}^S
\left(\hat{p}_{\bt s}\right)^{f_{\bt s}}, \\
P_2 &= \prod_{\bt\in\mathcal{S}^K}\prod_{s=1}^S
\left(\hat{p}_{\bt s}^L\right)^{f_{\bt s}^L}
\left(\hat{p}_{\bt s}^R\right)^{f_{\bt s}^R},
\end{aligned}
\end{equation}
respectively.  Because we have more free parameters to fit the
observed sequence statistics in the two-segment model, $P_2 \geq P_1$.
The generalized Jensen-Shannon divergence, a symmetric
variant of the relative entropy known more commonly as the
\emph{Kullback-Leibler divergence}, is then given by
\begin{equation}\label{equation:JensenShannondivergence}
\Delta(i) = \log\frac{P_2}{P_1} = 
\sum_{\mathbf{t}\in\mathcal{S}^K}\sum_{s = 1}^S \left[
-f_{\mathbf{t}s} \log \hat{p}_{\mathbf{t}s} +
f_{\mathbf{t}s}^L \log \hat{p}_{\mathbf{t}s}^L +
f_{\mathbf{t}s}^R \log \hat{p}_{\mathbf{t}s}^R \right].
\end{equation}
This test statistic, generalized from the Jensen-Shannon divergence
described in
Ref.~\citeonline{Lin1991IEEETransactionsonInformationTheory37p145},
measures quantitatively how much better the two-segment model fits
$\bx$ compared to the one-segment model.

\subsection{Paired Sliding Windows Segmentation Scheme}
\label{section:pairedslidingwindowssegmentationscheme}

A standard criticism on using sliding windows to detect segment
structure within a heterogeneous sequence is the compromise between
precision and statistical significance.  For the comparison between
two windowed statistics to be significant, we want the window size $n$
to be large.  On the other hand, to be able to determine a change
point precisely, we want the window size $n$ to be small.  There is
therefore no way, with a single window of length $n$, to independently
select both a desired statistical significance and desired precision.

In this appendix, we devise a sliding window segmentation scheme in
which, instead of one window, we use a pair of adjoining windows, each
of length $n$.  By comparing the left windowed statistics to the right
windowed statistics, a change point is detected at the center of the
pair of windows \emph{when} the two windowed statistics are most
different.  A given difference between the two windowed statistics
becomes more signficant as the window size $n$ is increased.  A larger
window size also suppresses statistical fluctuations, making it easier
to locate the change point.  Therefore, increasing the window size $n$
improves both statistical significance and precision, even though they
cannot be adjusted independently.

In App.~\ref{subsection:statisticswithinapairofslidingwindows}, we
describe the proper test statistic to use for change point detection
within the model selection framework.  Then in
App.~\ref{subsection:hypothesistestingwithapairofslidingwindows}, we
show how a similar test statistic spectrum can be obtained within the
hypothesis testing framework.  In
App.~\ref{subsection:performanceonrealgenomicsequences}, we show some
examples of the scheme being applied to real genomic sequences.  In
App.~\ref{subsection:meanfieldlineshape}, we derived the mean-field
lineshape of a domain wall in this paired sliding window segmentation
scheme, and use it to perform match filtering.

\subsubsection{Model Selection Within a Pair of Sliding Windows}
\label{subsection:statisticswithinapairofslidingwindows}

\begin{figure}[htbp]
\centering
\includegraphics[scale=0.8]{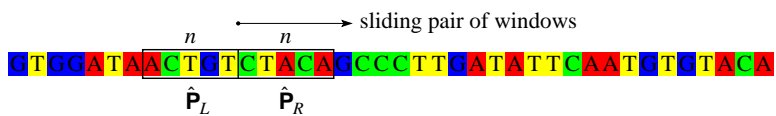}
\caption{A pair of sliding windows, each of length $n$.  A change
point at the center of the pair of sliding windows can be detected by
comparing the statistics within the left and right windows.}
\label{figure:pairedslidingwindow}
\end{figure}

To detect domain walls between different segments within a
heterogeneous sequence, we can slide a pair of adjoining windows each
of length $n$ across the sequence, and monitor the left and right
windowed statistics at different sequence positions, as shown in
Figure \ref{figure:pairedslidingwindow}.

If we model the different segments by Markov chains of order $K$, the
left and right windowed statistics are summarized by the transition
count matrices
\begin{equation}
\bF^L = \left[ f_{\mathbf{t}s}^L \right], \quad
\bF^R = \left[ f_{\mathbf{t}s}^R \right]
\end{equation}
respectively, where the transition counts sums to the window size,
\begin{equation}
\sum_{\mathbf{t}}\sum_s f_{\mathbf{t}s}^L = 
\sum_{\mathbf{t}}\sum_s f_{\mathbf{t}s}^R = n.
\end{equation}
From these transition count matrices, we can determine the
maximum-likelihood estimates
\begin{equation}
\hat{\bP}^L = \left[ p_{\mathbf{t}s}^L \right], \quad
p_{\mathbf{t}s}^L = \frac{f_{\mathbf{t}s}^L}{\sum_{s'}
f_{\mathbf{t}s'}^L}; \quad
\hat{\bP}^R = \left[ p_{\mathbf{t}s}^R \right], \quad
p_{\mathbf{t}s}^R = \frac{f_{\mathbf{t}s}^R}{\sum_{s'}
f_{\mathbf{t}s'}^R}
\end{equation} 
of the transition matrices for the left and right windows.

We then compute the transition count matrix
\begin{equation}
\bF = \left[ f_{\mathbf{t}s} = f_{\mathbf{t}s}^L + f_{\mathbf{t}s}^R \right],
\end{equation}
and therefrom the transition matrix
\begin{equation}
\hat{\bP} = \left[ p_{\mathbf{t}s} \right], \quad
p_{\mathbf{t}s} = \frac{f_{\mathbf{t}s}}{\sum_{s'} f_{\mathbf{t}s'}},
\end{equation}
assuming a one-segment model for the combined window of length $2n$,
before calculating the windowed Jensen-Shannon divergence using Eq.
\eqref{equation:JensenShannondivergence} in
App.~\ref{section:generalizedJensenShannondivergences}.  By sliding
the pair of windows along the sequence, we obtain a windowed
Jensen-Shannon divergence spectrum $\Delta(i)$, which tells us where
along the sequence the most statistically significant change points
are located.

\subsubsection{Hypothesis Testing With a Pair of Sliding Windows}
\label{subsection:hypothesistestingwithapairofslidingwindows}

Change point detection using statistics within the pair of sliding
windows can also be done within the hypothesis testing framework.
Within this framework, we ask how likely it is to find
maximum-likelihood estimates $\hat{\bP}^L$ for the left window, and
$\hat{\bP}^R$ for the right window, when the pair of windows straddles
a statistically stationary region generated by the transition matrix
$\bP$.

In the central limit regime, Whittle showed that the probability of
obtaining a maximum-likelihood estimate $\hat{\bP}$ from a finite
sequence generated by the transition matrix $\bP$ is given by
\cite{Whittle1955JRoyalStatSoc17p235}
\begin{equation}\label{equation:whittleformula}
P(\hat{\bP}|\bP) = C \exp\left[
\frac{1}{2} \sum_{\mathbf{t}}\sum_s\sum_{s'} \frac{n}{P_{\mathbf{t}}} 
\left(1 - \frac{\delta_{ss'}}{p_{\mathbf{t}s}}\right)
\left(\hat{p}_{\mathbf{t}s} - p_{\mathbf{t}s}\right)
\left(\hat{p}_{\mathbf{t}s'} - p_{\mathbf{t}s'}\right)
\right],
\end{equation}
where $C$ is a normalization constant, $n$ the length of the sequence,
and $P_{\mathbf{t}}$ is the equilibrium distribution of $K$-mers in
the Markov chain.

For $n \gg K$, the left and right window statistics are essentially
independent, and so the probability of finding $\hat{\bP}^L$ in the
left window and finding $\hat{\bP}^R$ in the right window, when the
true transition matrix is $\bP$, is $P(\hat{\bP}^L | \bP)
P(\hat{\bP}^R | \bP)$.  In principle we do not know what $\bP$ is, so
we replace it by $\hat{\bP}$, the maximum-likelihood transition matrix
estimated from the combined statistics in the left and right windows.
Based on Eq. \eqref{equation:whittleformula}, the test statistic
that we compute as we slide the pair of windows along the sequence is
the \emph{square deviation}
\begin{equation}\label{equation:centrallimitsquaredeviation}
r = -\sum_{\mathbf{t}}\sum_s\sum_{s'}
\frac{n}{\hat{P}_{\mathbf{t}}} 
\left(1 - \frac{\delta_{ss'}}{\hat{p}_{\mathbf{t}s}}\right)
\left[
\left(\hat{p}_{\mathbf{t}s}^L - \hat{p}_{\mathbf{t}s}\right)
\left(\hat{p}_{\mathbf{t}s'}^L - \hat{p}_{\mathbf{t}s'}\right) +
\left(\hat{p}_{\mathbf{t}s}^R - \hat{p}_{\mathbf{t}s}\right)
\left(\hat{p}_{\mathbf{t}s'}^R - \hat{p}_{\mathbf{t}s'}\right)
\right],
\end{equation}
which is more or less the negative logarithm of $P(\hat{\bP}^L | \bP)
P(\hat{\bP}^R | \bP)$.  To compare the square deviation spectrum
$r(i)$ obtained for different window sizes, we simply divide $r(i)$ by
the window size $n$.  From Eq.
\eqref{equation:centrallimitsquaredeviation}, we find that $r$ receive
disproportionate contributions from rare states
($\hat{P}_{\mathbf{t}}$ small) as well as rare transitions
($\hat{p}_{\mathbf{t}s}$ small).

\subsubsection{Application to Real Genomic Sequences}
\label{subsection:performanceonrealgenomicsequences}

\begin{figure}[hbtp]
\centering
\includegraphics[scale=0.5,clip=true]{EcoliK12.q.rwJSK0n1000.0.40k.eps}

\includegraphics[scale=0.5,clip=true]{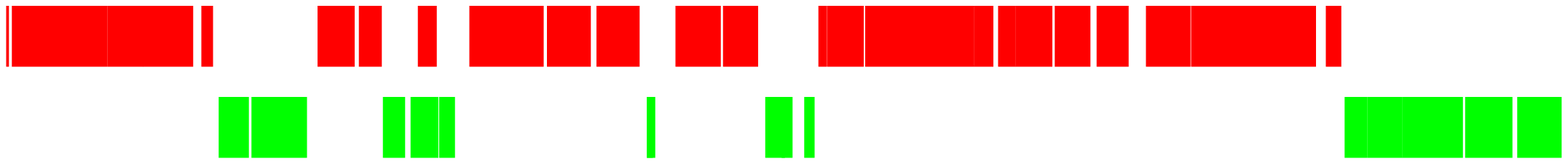}
\caption{The windowed $K = 0$ Jensen-Shannon divergence (magenta) and
square deviation (black) spectra in the interval $(0, 40000)$ of the
\emph{Escherichia coli} K-12 MG1655 genome, which has a length $N =
4639675$ bp.  Annotated genes on the positive (red) and negative
(green) strands are shown below the graph.}
\label{figure:EcoliK12qrwJSK0n1000i0i40k}
\end{figure}

The average length of coding genes in \emph{Escherichia coli} K-12
MG1655 is 948.9 bp.  This sets a `natural' window size to use for our
sliding window analysis.  In Figure
\ref{figure:EcoliK12qrwJSK0n1000i0i40k}, we show the windowed $K = 0$
Jensen-Shannon divergence and square deviation spectra for
\emph{Escherichia coli} K-12 MG1655, obtained for a window size of $n
= 1000$ bp, overlaid onto the distribution of genes.  As we can see
from the figure, the two spectra are qualitatively very similar, with
peak positions that are strongly correlated with gene and operon
boundaries \cite{Salgado2006NucleicAcidsResearch34pD394}.

For example, we see that the strongest peak in the $n = 1000$ windowed
spectrum is at $i \sim 30000$.  The gene \emph{dapB}, believed to be
an enzyme involved in lysine (which consists solely of purines)
biosynthesis, lies upstream of this peak, while the \emph{carAB}
operon, believed to code for enzymes involved in pyrimidine
ribonucleotide biosynthesis, lies downstream of the peak.  Another
strong peak marks the end of the \emph{carAB} operon, distinguishing
it statistically from the gene \emph{caiF}, and yet another strong
peak distinguishes \emph{caiF} from the \emph{caiTABCDE} operon, whose
products are involved in the central intermediary metabolic pathways,
further downstream.

\begin{figure}[hbtp]
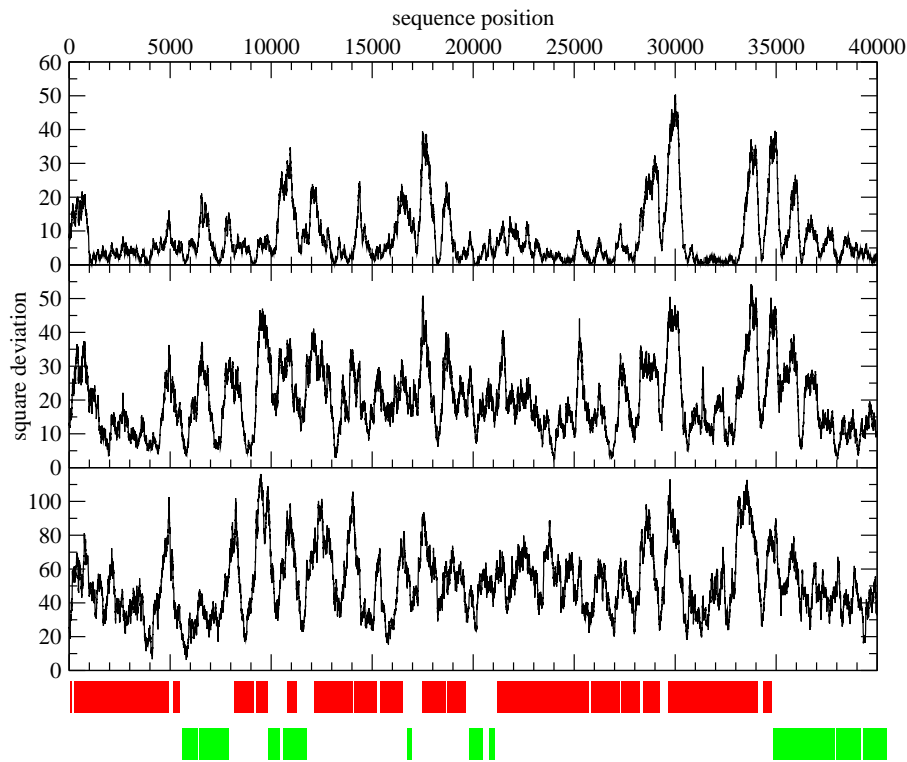

\centering
\includegraphics[scale=0.5,clip=true]{EcoliK12.q.rK0K1K2n1000.0.40k.eps}

\includegraphics[scale=0.5,clip=true]{NC_000913.gene.layout.0.40k.eps}
\caption{The windowed $K = 0$ (top), $K = 1$ (middle), and $K = 2$
(bottom) square deviation spectra in the interval $(0, 40000)$ of the
\emph{E. coli} K-12 MG1655 genome, which has a length of $N = 4639675$
bp.  Annotated genes on the positive (red) and negative (green)
strands are shown below the graph.}
\label{figure:EcoliK12qrK0K1K2n1000i0i40k}
\end{figure}

In Figure \ref{figure:EcoliK12qrK0K1K2n1000i0i40k}, we show the square
deviation spectra for the same $(0, 40000)$ interval of the \emph{E.
coli} K-12 MG1655 genome, but for different Markov-chain orders $K =
0, 1, 2$.  As we can see, these square deviation spectra share many
qualitative features, but there are also important qualitative
differences.  For example, the genes \emph{talB} and \emph{mogA},
which lies within the interval $(8200, 9900)$, are not strongly
distinguished from the genes \emph{yaaJ} upstream and \emph{yaaH}
downstream at the 1-mer ($K = 0$) level.  They are, however, strongly
distinguished from the flanking genes at the 2-mer ($K = 1$) and 3-mer
($K = 2$) levels.

\subsubsection{Mean-Field Lineshape and Match Filtering}
\label{subsection:meanfieldlineshape}

In the second situation shown in Fig.~\ref{figure:windowedcases}, let
us label the two mean-field segments $a$ and $b$, with lengths $N_a$
and $N_b$.  Suppose it is the left window that straddles both $a$ and
$b$, while the right window lies entirely within $b$.  The
right-window counts are then simply
\begin{equation}
f_{\mathbf{t}s}^R = \frac{n}{N_b}\, f_{\mathbf{t}s}^b,
\end{equation}
while the left-window counts contain contributions from both $a$ and
$b$, i.e.
\begin{equation}
f_{\mathbf{t}s}^L = \frac{n - z}{N_a}\, f_{\mathbf{t}s}^a + 
\frac{z}{N_b}\, f_{\mathbf{t}s}^b,
\end{equation}
where $z$ is the distance of the domain wall from the center of the
pair of windows.  The total counts from both windows are then
\begin{equation}
f_{\mathbf{t}s} = \frac{n - z}{N_a}\, f_{\mathbf{t}s}^a + 
\frac{z}{N_b}\, f_{\mathbf{t}s}^b + \frac{n}{N_b}\, f_{\mathbf{t}s}^b.
\end{equation}

Using the transition counts $f_{\mathbf{t}s}^L$, $f_{\mathbf{t}s}^R$,
and $f_{\mathbf{t}s}$, we then compute the maximum-likelihood
transition probabilities $\hat{p}_{\mathbf{t}s}^L$,
$\hat{p}_{\mathbf{t}s}^R$, and $\hat{p}_{\mathbf{t}s}$, before
substituting the transition counts and transition probabilities into
Eq.~\eqref{equation:JensenShannondivergence} for the Jensen-Shannon
divergence.  Because of the logarithms in the definition for the
Jensen-Shannon divergence, we get a complicated function in terms of
the observed statistics $f_{\mathbf{t}s}^a$, $f_{\mathbf{t}s}^b$,
$N_a$ and $N_b$, and the distance $z$ between the domain wall and the
center of the pair of windows.  Different observed statistics
$f_{\mathbf{t}s}^a$, $f_{\mathbf{t}s}^b$, $N_a$ and $N_b$ give
mean-field divergence functions of $z$ that are not related by a
simple scaling.  However, these mean-field divergence functions
$\Delta(z)$ do have qualitative features in common:
\begin{enumerate}

\item $\Delta(z) = 0$ for $|z| \geq n$, where the pair of windows is
entirely within $a$ or entirely within $b$;

\item $\Delta(z)$ is maximum at $z = 0$, when the center of the pair
of windows coincide with the domain wall;

\item $\Delta(z)$ is convex everywhere within $|z| < n$, except at $z
= 0$.

\end{enumerate}
This tells us that the position and strength of the domain wall
between two mean-field segments both longer than the window size $n$
can be determined exactly.

\begin{figure}[htbp]
\centering
\includegraphics[scale=0.45,clip=true]{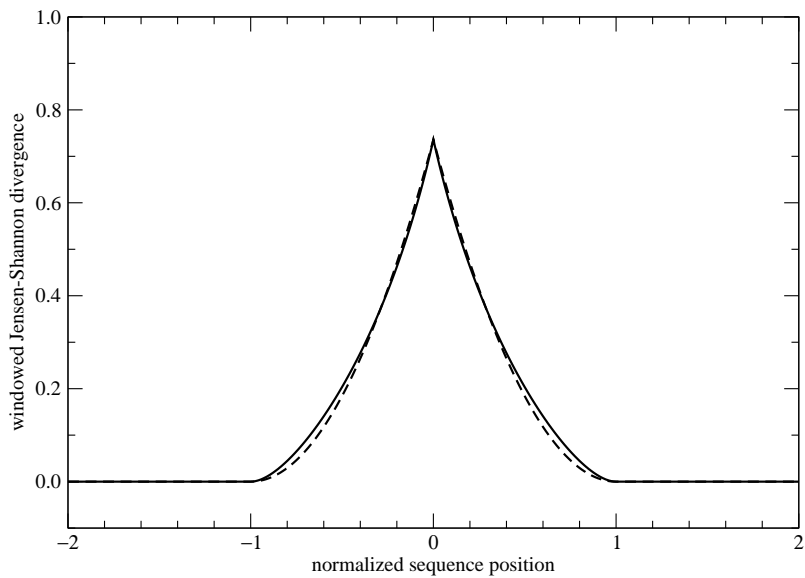}
\caption{The Jensen-Shannon divergence $\Delta(z)$ (solid curve) of a
pair of sliding windows of length $n = 1$ as a function of the
distance $z$ between the domain wall separating a mean-field binary
segment $a$ with $P_a(0) = 1 - P_a(1) = 0.9$ and a mean-field binary
segment $b$ with $P_b(0) = 1 - P_b(1) = 0.1$, and the center of the
pair of windows.  Also shown as the dashed curve is a piecewise
quadratic function which rises from $z = \pm 1$ to the same maximum at
$z = 0$, but vanishes everywhere else.}
\label{figure:wJSlineshape}
\end{figure}

In Figure \ref{figure:wJSlineshape} we show $\Delta(z)$ for two binary
$K = 0$ mean-field segments, where $P_a(0) = 1 - P_a(1) = 0.9$, and
$P_b(0) = 1 - P_b(1) = 0.1$.  We call the peak function $\Delta(z)$
the \emph{mean-field lineshape} of the domain wall.  As we can see
from Figure \ref{figure:wJSlineshape}, this mean-field lineshape can
be very well approximated by the piecewise quadratic function
\begin{equation}\label{equation:meanfieldJS}
\tilde{\Delta}(z) = \begin{cases}
\left(1 + \frac{z}{n}\right)^2 \bar{\Delta}(0), & -1 < z < 0; \\
\left(1 - \frac{z}{n}\right)^2 \bar{\Delta}(0), & 0 \leq z < 1; \\
0, & \text{everywhere else}, \end{cases}
\end{equation}
where $\bar{\Delta}(0)$ is the mean-field Jensen-Shannon divergence of
the domain wall at $z = 0$.  If instead of the windowed Jensen-Shannon
divergence $\Delta(z)$, we compute the windowed square deviation
$r(z)$ in the vicinity of a domain wall, we will obtain a mean-field
lineshape that is strictly piecewise quadratic, i.e.
\begin{equation}\label{equation:meanfieldr}
\tilde{r}(z) = \begin{cases}
\left(1 + \frac{z}{n}\right)^2 \bar{r}(0), & -1 < z < 0; \\
\left(1 - \frac{z}{n}\right)^2 \bar{r}(0), & 0 \leq z < 1; \\
0, & \text{everywhere else}, \end{cases}
\end{equation}
where $\bar{r}(0)$ is the mean-field square deviation of the domain
wall at $z = 0$.

\begin{figure}[hbtp]
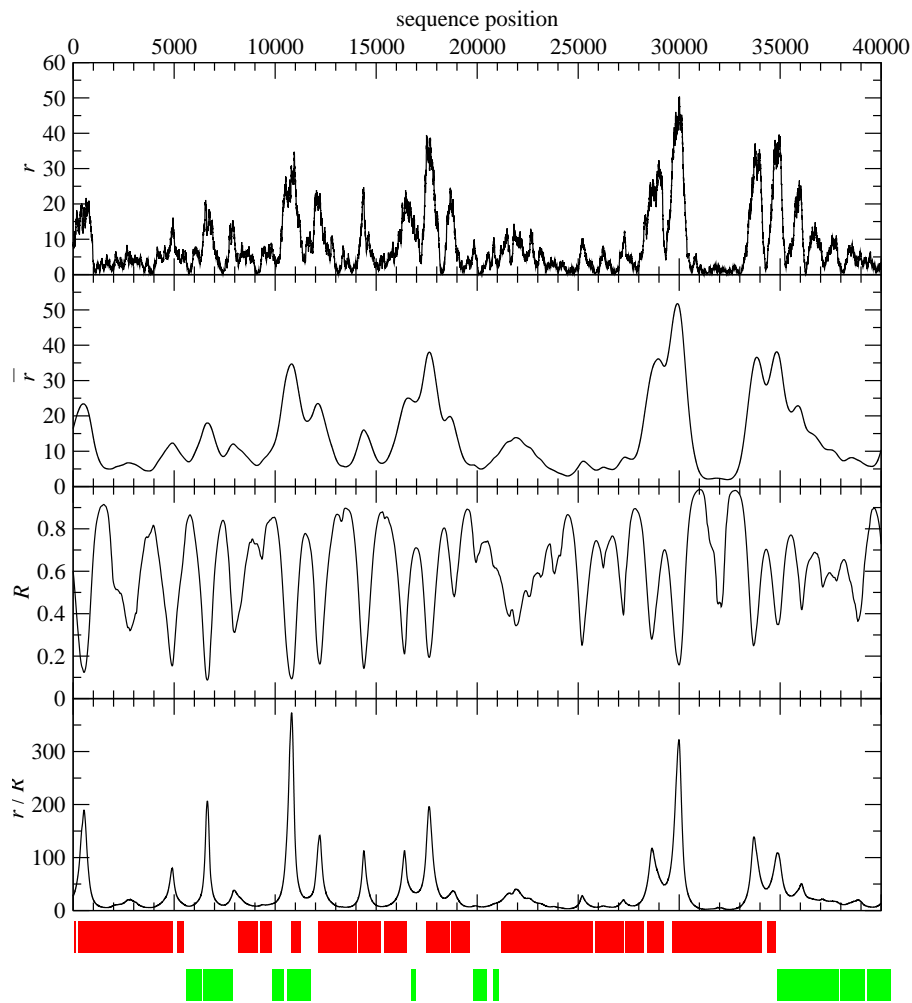

\centering
\includegraphics[scale=0.5,clip=true]{EcoliK12.q.rrmrmRK0n1000.0.40k.eps}

\includegraphics[scale=0.5,clip=true]{NC_000913.gene.layout.0.40k.eps}
\caption{The interval $0 \leq i \leq 40000$ of the \emph{E. coli} K-12
MG1655 genome ($N = 4639675$ bp), showing (top to bottom) the windowed
$K = 0$ square deviation spectrum $r(i)$, the match-filtered square
deviation spectrum $\bar{r}(i)$, the residue spectrum $R(i)$, and the
quality enhanced square deviation spectrum $\bar{r}(i)/R(i)$.
Annotated genes on the positive (red) and negative (green) strands are
shown below the graph.}
\label{figure:EcoliK12qrrmrmRK0n1000i0i40k}
\end{figure}

Going back to a real sequence composed of two nearly stationary
segments of discrete bases, we expect to find statistical fluctuations
masking the mean-field lineshape.  But now that we know the mean-field
lineshape is piecewise quadratic for the square deviation $r(z)$ (or
very nearly so, in the case of the windowed Jensen-Shannon divergence
$\Delta(z)$), we can make use of this piecewise quadratic mean-field
lineshape to match filter the raw square deviation spectrum.  We do
this by assuming that there is a mean-field square-deviation peak at
each sequence position $i$, fit the spectrum within $(i - n, i + n)$
to the mean-field lineshape in Eq. \eqref{equation:meanfieldr}, and
determine the smoothed spectrum $\bar{r}(i)$.  In Fig. 
\ref{figure:EcoliK12qrrmrmRK0n1000i0i40k}, we show the match-filtered
square deviation spectrum $\bar{r}(i)$ in the interval $0 \leq i \leq
40000$ of the \emph{E. coli} K-12 MG1655 genome.  As we can see,
$\bar{r}(i)$ is smoother than $r(i)$, but the peaks in $\bar{r}(i)$
are also so broad that distinct peaks in $r(i)$ are not properly
resolved.

Fortunately, more information is available from the match filtering.
We can also compute how well the raw spectrum $r(j)$ in the interval
$i - n \leq j \leq i + n$ match the mean-field lineshape
$\tilde{r}(j)$ by computing the residue
\begin{equation}
R(i) = \sum_{j = i - n}^{i + n} \left[r(j) - \tilde{r}(j)\right]^2.
\end{equation}
filtering the raw divergence spectrum.  In Fig.
\ref{figure:EcoliK12qrrmrmRK0n1000i0i40k}, we show the residue
spectrum $R(i)$ for the $0 \leq i \leq 40000$ region of the \emph{E.
coli} K-12 MG1655 genome.  In the residue spectrum, we see a series of
dips at the positions of peaks in the square deviation spectrum.
Since $R(i)$ is small when the match is good, and large when the match
is poor, $1/R(i)$ can be thought of as the quality factor of a square
deviation peak.  A smoothed, and accentuated spectrum is obtained when
we divide the smoothed square deviation by the residue at each point.
The quality enhanced square deviation spectrum $\bar{r}(i)/R(i)$ is
also shown in Fig. \ref{figure:EcoliK12qrrmrmRK0n1000i0i40k}.  It is
much more convenient to determine the position of significant domain
walls from such a spectrum.

\end{appendix}

\end{document}